\documentclass[11pt]{article}
\usepackage{graphics}
\usepackage{epsfig}

\def\DN{\Delta N_\nu}
\def\be{\begin{equation}}  
\def\bea{\begin{eqnarray}}  
\def\ee{\end{equation}}     
\def\eea{\end{eqnarray}}     
\def\gsim{\mathrel{\raise.3ex\hbox{$>$\kern-.75em\lower1ex\hbox{$\sim$}}}}
\def\lsim{\mathrel{\raise.3ex\hbox{$<$\kern-.75em\lower1ex\hbox{$\sim$}}}}

\begin{document}

\vskip .7cm

\vspace{30cm}

\begin{center}

{\Large \bf  Cosmology with mirror dark matter I: \\
\vspace{0.3cm}
linear evolution of perturbations }

\vskip .7cm

{\large Paolo Ciarcelluti}

\vskip .7cm

{\it Dipartimento di Fisica, Universit\`a di L'Aquila, 67010 Coppito  
AQ, and  \\  INFN, Laboratori Nazionali del Gran Sasso, 67010 Assergi AQ, 
Italy \\ 
E-mail: {\tt ciarcelluti@lngs.infn.it}
\vspace{8pt}
}

\end{center}

\vskip .3cm

\begin{abstract}

This is the first paper of a series devoted to the study of the 
cosmological implications of the parallel mirror world with the same 
microphysics as the ordinary one, but having smaller temperature, 
with a limit set by the BBN constraints.
The difference in temperature of the ordinary and mirror sectors generates 
shifts in the key epochs for structure formation, which proceeds in the 
mirror sector under different conditions.
We consider adiabatic scalar primordial perturbations as an input and 
analyze the trends of all the relevant scales for structure formation 
(Jeans length and mass, Silk scale, horizon scale) for both ordinary 
and mirror sectors, comparing them with the CDM case. 
These scales are functions of the fundamental parameters of the theory
(the temperature of the mirror plasma and the amount of mirror baryonic 
matter), and in particular they are influenced by the difference between the 
cosmological key epochs in the two sectors.
Then we used a numerical code to compute the evolution in linear 
regime of density perturbations for all the components of a Mirror Universe: 
ordinary baryons and photons, mirror baryons and photons, 
and possibly cold dark matter.
We analyzed the evolution of the perturbations for different values of 
mirror temperature and baryonic density, and obtained that 
for $ x = T' / T $ less than a typical value $ x_{\rm eq} $, 
for which the mirror baryon-photon decoupling happens before the 
matter-radiation equality, mirror baryons are equivalent to the CDM 
for the linear structure formation process.
Indeed, the smaller the value of $x$, the closer mirror dark matter 
resembles standard cold dark matter during the linear regime. 

\end{abstract}

\vspace{0.3cm}


\section{Introduction}

The present cosmological observations strongly support the main 
predictions of the inflationary scenario: first, the Universe is flat, with the 
total energy density very close to the critical $\Omega_0 \approx 1$, 
and second, primordial density perturbations have nearly flat spectrum, 
with the scalar spectral index $n \approx 1$. 
The non-relativistic matter gives only a small fraction of the present 
energy density, about $ \Omega_m \simeq 0.25 $, while the rest is 
attributed to the vacuum energy (cosmological term or dark energy) 
$ \Omega_\Lambda \simeq 0.75 $. 
The fact that $ \Omega_m $ and $ \Omega_\Lambda $ are of the same 
order gives rise to the so called cosmological coincidence problem: 
why we live in an epoch when $ \rho_m \sim \rho_\Lambda $, if in the 
early Universe one had $ \rho_m \gg \rho_\Lambda $ and in the 
late Universe one would expect $ \rho_m \ll \rho_\Lambda $? 
The answer can be only related to an anthropic principle: 
the matter and vacuum energy densities scale differently with the 
expansion of the Universe, $ \rho_m \propto a^{-3} $ and 
$ \rho_\Lambda \propto $ const., hence they have to coincide at some 
moment, and we are just happy to be here. 
Moreover, for substantially larger $ \rho_\Lambda $ no galaxies could be 
formed and thus there would not be anyone to ask this question. 

On the other hand, the matter in the Universe has two components, 
visible and dark: $ \Omega_m = \Omega_b +\Omega_d $.  
The visible matter consists of baryons  with $ \Omega_b \approx 0.04 $, 
while the dark matter with $ \Omega_d \approx 0.2 $ is constituted by some 
hypothetic particle species very weakly interacting with the observable 
matter. 
It is a tantalizing question why the visible and dark 
components have so close energy densities?  
Clearly, the ratio $ \rho_d / \rho_b $
does not depend on time as far as with the expansion of the 
Universe both $ \rho_b $ and $ \rho_d $ scale as $ \propto a^{-3} $. 

In view of the standard cosmological paradigm, there is no good reason 
for having $ \Omega_d \sim \Omega_b $, as far as the visible and dark 
components have different origins. 
The density of the visible matter is $ \rho_b = M_N n_b $, where 
$ M_N \simeq 1 $ GeV is the nucleon mass, and $ n_b $ is the baryon 
number density of the Universe.   
The latter should be produced in a very early Universe 
by some baryogenesis mechanism, which is presumably related 
to some B and CP-violating physics at very high energies. 
The baryon number per photon $\eta=n_b/n_\gamma$ is very small. 
Observational data on the primordial abundances of light elements and the 
recent results on the cosmic microwave background radiation (CMB) 
anisotropies nicely converge to the value $\eta \approx 6\times 10^{-10}$.

As for dark matter, it is presumably constituted by some cold relics with 
mass $ M $ and number density $ n_d $, and $ \rho_d = M n_d $. 
The most popular candidate for cold dark matter (CDM) is provided by the 
lightest supersymmetric particle (LSP) with $ M_{LSP} \sim 1 $ TeV, and 
its number density $ n_{LSP} $ is fixed by its annihilation cross-section. 
Hence $ \rho_b \sim \rho_{LSP} $ requires that 
$ n_b / n_{LSP} \sim M_{LSP} / M_N $ and the origin of such a conspiracy 
between four independent parameters is unclear: 
the value $ M_N $ is fixed by the QCD scale while $ M_{LSP} $ is related 
to the supersymmetry breaking scale, $ n_b $ is determined by B and CP 
violating properties of the particle theory at very high energies whereas  
$ n_{LSP} $ strongly depends on the supersymmetry breaking details. 
Within the parameter space of the MSSM (Minimal Supersymmetric 
Standard Model) it could vary within several orders of magnitude, and 
moreover, in either case it has nothing to do with the B and CP 
violating effects.  
The situation looks even more obscure if the dark component is related 
e.g. to the primordial oscillations of a classic axion field, in which case 
the dark matter particles constituted by axions are superlight, with mass 
$ \ll 1 $ eV, but they have a condensate with enormously high number 
density.

A new sight on the problem of dark matter can be given by the concept 
of mirror world. 
This idea was studied in ref. \cite{mirror} and then its phenomenological 
and cosmological implications discussed in several later papers.
In particular, after its first applications to non-baryonic dark matter 
\cite{blinkhlo}, the mirror matter hypothesis has attracted a significant 
interest over the last years and has been invoked in 
many physical and astrophysical questions: 
large scale structure of the Universe \cite{ignavol-lss,paolo}, 
galactic halo \cite{mir_halo},
MACHOs \cite{Macho}, 
gamma ray bursts \cite{mir_GRB}, 
flavour and CP violation \cite{assione}
orthopositronium lifetime \cite{ortho,bader}, 
interpretation of dark matter detection experiments \cite{mir_dama}, 
meteoritic event anomalies \cite{mir_meteor,detect_mir_frag}, 
close-in extrasolar planets \cite{mir_planet}, 
Pioneer spacecraft anomalies \cite{pioneer}.

Mirror world can be considered as a hidden sector of particles and 
interactions which are exactly the same as in our visible world\footnote{
 The mirror parity could be spontaneously broken and the weak 
interaction scales $ \langle \phi \rangle =v $ and $ \langle \phi' \rangle =v' $ 
could be different, which leads to somewhat different particle physics in the 
mirror sector \cite{broken}. 
In this paper we treat only the simplest case, $ v = v' $, in which the mirror 
sector has exactly the same physics as the ordinary one.}, 
and coupling to our world essentially by gravity.
Besides the gravity, the two sectors could communicate by other means. 
In particular, ordinary photons could have kinetic mixing with mirror 
photons \cite{mixing}, or ordinary (active) neutrinos could mix with mirror 
(sterile) neutrinos \cite{neutrino}.

If the mirror sector exists, then the Universe along with the ordinary (O) 
photons, neutrinos, baryons, etc. should contain their mirror (M) partners.  
One could naively think that due to mirror parity the ordinary and mirror 
particles should have the same cosmological abundances and hence the O 
and M sectors should have the same cosmological evolution. 
However, this would be in conflict with the Big Bang nucleosynthesis 
(BBN) bounds on  the effective number of extra light neutrinos, since the 
mirror photons, electrons and neutrinos would give a contribution to 
the Hubble expansion rate equivalent to $\DN\simeq 6.14$ \cite{inflation2}.
Therefore, even if their microphysics is the same, initial conditions are not 
the same in both sectors, and in the early Universe the M system should 
have a lower temperature than ordinary particles. 
This situation is plausible if the two systems are born with different 
temperatures (the post-inflationary reheating temperature in the M sector is 
lower than in the visible one \cite{inflation2,inflation}), they interact very 
weakly,  so that they do not come into thermal equilibrium 
with each other (condition automatically fulfilled  if the two worlds 
communicate only via gravity), and both of them expand adiabatically.
So they evolve independently and their temperatures remain different 
at later stages.

At present, the temperature of ordinary relic photons is $T\approx 2.72$ K, 
and the mass density of ordinary baryons constitutes about $5\%$ of the 
critical density. 
Mirror photons should have smaller temperature $ T' < T $, so their 
number density is $ n'_\gamma = x^3 n_\gamma $, where $ x=T'/T $.
\footnote{
From now on all quantities of the mirror sector will be marked by $'$ to 
distinguish from the ones belonging to the observable or ordinary world.}
This ratio is a key parameter in our further considerations since it remains 
nearly invariant during the expansion of the Universe. 
The BBN limit on $\DN$ implies the upper bound $x < 0.64\, \DN^{1/4}$. 
If we assume $ \DN \lsim 1 $, we obtain $ x \lsim 0.64 $ as a limit 
for the temperature of the mirror sector \cite{bcv}.
The dependence of $ x $ on $\DN$ is indeed very weak, so 
even a value $ x \lsim 0.7 $ could be compatible with the current 
experimental status.
As for mirror baryons, their number density $n'_b$  can be larger than 
$n_b$, and if the ratio $\beta = n'_b/n_b$ is about 5 or so, they could 
constitute the dark matter of the Universe. 

In this view, the mirror world idea could give a new twist to dark 
matter problem. 
Once the visible matter is built up by ordinary baryons, then the mirror 
baryons could constitute dark matter in a natural way.  
They interact with mirror photons, however they are dark 
in terms of the ordinary photons. 
The mass of M baryons is the same as the ordinary 
one, $ M_N' = M_N $, and so we have $ \beta = \Omega'_b / \Omega_b $.
In addition, as far as the two sectors have the same particle physics, it is 
natural to think that the M baryon number density $n'_b$ is determined by 
the baryogenesis mechanism which is similar to the one which fixes the 
O baryon density $n_b$. 
Thus, one could question whether the ratio $ \beta $ could be naturally 
order 1 or somewhat bigger.     

There are several baryogenesis mechanisms as are GUT baryogenesis, 
leptogenesis, electroweak baryogenesis, etc. 
At present it is not possible to say definitely which of the known 
mechanisms is responsible for the observed baryon asymmetry in the 
ordinary world. 
However, it is most likely that the baryon asymmetry in the mirror world 
is produced by the same mechanism and moreover, the properties of the 
$B$ and CP violation processes are parametrically the same in both cases.
But the mirror sector has a lower temperature than ordinary one, 
and so at epochs relevant for baryogenesis the out-of-equilibrium 
conditions should be easier fulfilled for the M sector.

In particular, we know that in certain baryogenesis scenarios 
the M world gets a larger baryon asymmetry than the O sector, 
and it is pretty plausible that $\beta \gsim 1$ \cite{bcv}.
This situation emerges in a particularly appealing way in the leptogenesis 
scenario due to the lepton number  leaking from the O to the M sector 
which leads to  $n'_b \geq n_b$, and can thus explain the near 
coincidence of visible and dark components in a rather natural way 
\cite{baryo-lepto}.    

Given its consistency with all the experimental situation so far, 
it's important to construct a complete theory of cosmology with mirror 
dark matter and test this scenario. 
Thus the aim of this work (described in this paper and the next one of this 
series, hereafter referred to as Paper 2 \cite{paper2}) is contained in a 
question: ``is mirror matter still a reliable dark matter candidate?'' 
Its emergence arises also from the astrophysical problems encountered 
by the standard candidate, the cold dark matter (CDM), as for example 
the central galactic density profiles or the number of small satellites. 

In this first paper, we focus on the structure formation theory in linear 
regime, analyzing the trends of scales in both sectors and comparing them 
with the CDM case, and finally showing the evolution of perturbations in 
all the components of a Mirror Universe\footnote{
The expression ``Mirror Universe'' is clearly misleading, 
since one could think that there is another Universe, while it is just one, 
but made of two sectors. 
Nevertheless, this expression is sometimes used to shortly refer to this 
scenario.} 
(ordinary and mirror photons and baryons, and possibly CDM).
In doing so, we largely extend our previous studies of ref.~\cite{paolo}, 
analyzing the details of the structure formation and evolution in a mirror 
scenario. 
The paper is organized as follows.
In section 2 we show that, due to the temperature difference, in mirror sector 
the key epochs for structure formation occur at different redshifts than 
in the observable sector.
The section 3 analyzes the relevant scales for structure formation (sound 
speed, Jeans length and mass, Silk mass) as functions of the temperature 
and the baryonic density of the mirror sector.
The main differences with respect to ordinary baryonic and CDM scenarios 
are also discussed.
In section 4 we study the possible mirror scenarios for the growth of 
primordial perturbations.
Section 5 is devoted to the study of the temporal evolution of perturbations 
in all the ordinary and mirror components.
Finally, in section 6 we briefly summarize our findings.


\section{\bf  Mirror baryons as dark matter }
\label{mir_dm}

Since it interacts only gravitationally with our ordinary sector, 
the mirror matter is a {\sl natural dark} matter candidate. 
At present there are observational evidences that dark matter exists and 
it's density is about 5 times that of ordinary baryons, but in previous 
section we said that this is not a problem for mirror scenario, because 
the same microphysics does not imply the same initial conditions in 
both O and M sectors.

In the most general context, the present energy density contains relativistic 
(radiation) component $ \Omega_r $, non-relativistic (matter) component 
$ \Omega_m $ and the vacuum energy density $ \Omega_\Lambda $. 
According to the inflationary paradigm the Universe should be almost flat,  
$\Omega_0=\Omega_m + \Omega_r + \Omega_\Lambda \approx 1$, which 
well agrees with the results on the CMB anisotropy.

If we consider now a Mirror Universe, i.e. a Universe made of two sectors, 
$\Omega_{r}$ and $\Omega_{m}$ represent the total amount of 
radiation and matter of both ordinary and mirror sectors: 
$ \Omega_{r} = (\Omega_{r})_{\rm O} + (\Omega_{r})_{\rm M} $ and 
$ \Omega_{m} = (\Omega_{m})_{\rm O} + (\Omega_{m})_{\rm M} $. 
The two parameters describing the mirror sector are the ratio of the 
temperatures of two sectors $ x $ and the relative amount of mirror baryons 
compared to the ordinary ones $ \beta $:
\be \label{mir-par}
x = {T' \over T} \lsim 0.64
\;\;\;\;\;\;\;\;\;\;\;\;\;\;\;\;\;\; 
\beta = {\Omega'_b \over \Omega_b} \gsim 1 \;.
\ee
In the above expressions, the first limit comes from the BBN bound on the 
effective number of extra light neutrinos (see previous section) 
and the second one from the hypothesis that mirror baryonic contribution 
to dark matter has cosmological relevance.
In the context of our model, as explained in \cite{bcv}, the relativistic 
fraction is represented by the ordinary and mirror photons and neutrinos, 
and, using the expression for the ordinary degrees of freedom in a Mirror 
Universe, $ \bar g(T) = g_\ast (T) (1 + x^4) $, and the value of the observable 
radiation energy density 
$ (\Omega_r)_{\rm O} \, h^2 \simeq 4.2 {\times} 10^{-5} $, it is given by 
\be
\Omega_r = 4.2 \times 10^{-5}\,h^{-2}\,(1+x^4) \simeq 4.2 \times 10^{-5}\,h^{-2} \;,
\ee
where the contribution of the mirror species, expressed by the additional 
term $ x^4 $, is negligible in view of the BBN constraint $ x < 0.64 $. 
As for the non-relativistic component, it contains the O baryon fraction 
$\Omega_b$ and the M baryon fraction $\Omega'_b = \beta\Omega_b$,
while the other types of dark matter could also present.
Obviously, since mirror parity doubles {\sl all} the ordinary particles, even if 
they are ``dark'' (i.e., we are not able to detect them now), whatever the form 
of dark matter made by some exotic ordinary particles, there will exist a 
mirror partner made by the mirror counterpart of these particles. 
In the context of supersymmetry, the CDM component could exist in the form 
of the lightest supersymmetric particle (LSP).  
It is interesting to remark that the mass fractions of the ordinary and mirror 
LSP are related as $\Omega'_{LSP} \simeq x\Omega_{LSP}$. 
In addition, a significant HDM component $\Omega_\nu$ could be due to 
neutrinos with order eV mass. 
The contribution of the mirror massive neutrinos scales as 
$\Omega'_\nu = x^3 \Omega_\nu$ and thus it is irrelevant. 
In any case, considering the only CDM component, which is now the 
preferred candidate, we can combine both the ordinary and mirror 
components, since their physical effects are exactly the same. 
Thus we consider a matter composition of the Universe expressed in 
general by
\be
\Omega_m=\Omega_b+\Omega'_b+\Omega_{CDM}
                 = \Omega_b (1+\beta) + \Omega_{CDM}\;.
\ee
  
\begin{figure}[h]
  \begin{center}
    \leavevmode
    \epsfxsize = 10cm
    \epsffile{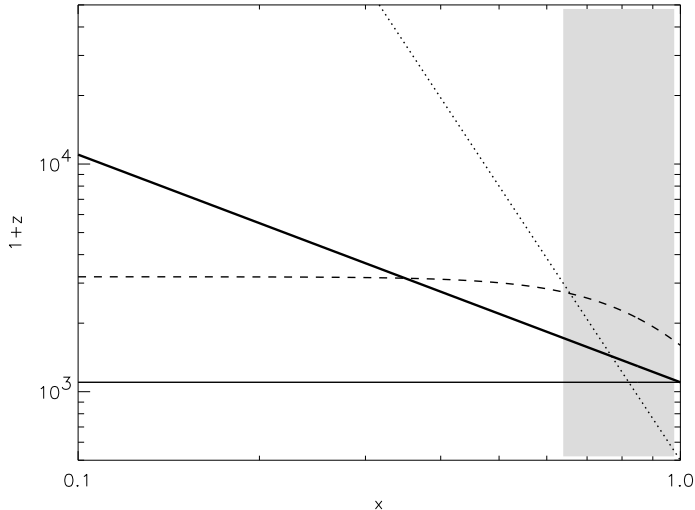}
  \end{center}
\vskip -5cm
\caption{\small 
The M photon decoupling redshift $1+z_{\rm dec}'$ as a function of $x$ 
(thick solid). The horizontal thin solid line marks the ordinary 
photon decoupling redshift $1+z_{\rm dec} = 1100$. 
We also show the matter-radiation equality redshift
$1+z_{\rm eq}$ (dash) and the mirror Jeans-horizon mass equality
redshift $1+z'_c$ (dot) for the case $\Omega_m h^2 =0.135$. 
The shaded area $x> 0.64$ is excluded by the BBN limits. }
\label{figzz}
\end{figure}

The important moments for the structure formation are related to the 
matter-radiation equality (MRE) epoch and to the plasma recombination and 
matter-radiation decoupling (MRD) epochs in both sectors. 
The MRE occurs at the redshift 
\be \label{z-eq} 
1+z_{\rm eq}= {{\Omega_m} \over {\Omega_r}} \approx 
 2.4\cdot 10^4 {{\Omega_{m}h^2} \over {1+x^4}} \;,~~
\ee
which is {\sl always smaller than the value obtained for an ordinary 
Universe}, but approximates it for low $x$ (see fig.~\ref{figzz}). 
If we consider only ordinary and mirror baryons and photons, we find
\be \label{z-eq_2} 
1+z_{\rm eq} = 
 {{\rho_b \left( 1+\beta \right)} \over {\rho_{\gamma } \left( 1+x^4 \right)}} 
 {\left( 1+z \right)} =
 {{\rho_b' \left( 1+\beta^{-1} \right)} \over {\rho_{\gamma }' \left( 1+x^{-4} \right)}} 
 {\left( 1+z \right)} \;,~~
\ee
where the baryon and photon densities refer to the redshift $z$.
This implies that, with the addition of a mirror sector, the matter-radiation 
equality epoch shifts toward earlier times as \footnote{
In the most general case, where there is also some other dark matter 
component, as the CDM, the parameter $\beta$ could be the sum of two 
terms, $ \beta + \beta_{\rm CDM} $.}
\be \label{shifteq} 
1+z_{\rm eq} ~~ \longrightarrow ~~ 
  { {\left( 1+\beta \right)} \over {\left( 1+x^4 \right)} } ~ (1+z_{\rm eq}) \;.
\ee

The MRD, instead, takes place in every sector only after the most of 
electrons and protons recombine into neutral hydrogen and the free electron 
number density $n_{e}$ diminishes, so that the interaction rate of the 
photons
$\Gamma_\gamma=n_{e}\sigma_{T}=X_{e}\eta n_{\gamma} \sigma_{T}$ 
drops below the Hubble expansion rate $H(T)$, where 
$\sigma_T$ is the Thomson cross section, $X_e=n_e/n_b$ is the fractional 
ionization, and $ \eta = n_b/n_\gamma $ is the baryon to photon ratio.
In condition of chemical equilibrium, $ X_e $ 
is given by the Saha equation, which for $ X_e \ll 1$ reads
\be \label{Saha} 
X_e \approx (1-Y_4)^{1/2}\; {{0.51} \over {\eta^{1/2}}} 
\left({T} \over {m_e}\right)^{-3/4} e^{-B/2T} \;,
\ee 
where $ B=13.6 $ eV is the hydrogen binding energy and $ Y_4 $ is the 
$^4$He abundance.
Thus we obtain the familiar result that in our Universe the MRD takes place 
in the matter domination period, at the temperature 
$T_{\rm dec} \simeq 0.26$ eV, which corresponds to redshift 
$1+z_{\rm dec}=T_{\rm dec}/T_0 \simeq 1100$.

The MRD temperature in the M sector $T'_{\rm dec}$ can be calculated 
following the same lines as in the ordinary one. 
Due to the fact that in either case the photon decoupling occurs when the 
exponential factor in eq.~(\ref{Saha}) becomes very small, we have 
$T'_{\rm dec} \simeq T_{\rm dec}$, up to small corrections related to 
$\eta'$, $Y'_4$ different from $\eta$, $Y_4$ 
(for more details see ref.~\cite{bcv}). 
Hence, considering that $T' = x \cdot T$, we obtain
\be \label{z'_dec}
1+z'_{\rm dec} \simeq x^{-1} (1+z_{\rm dec}) 
\simeq 1.1\cdot 10^3 x^{-1} \;, ~~
\ee
so that {\sl the MRD in the M sector occurs earlier than in the ordinary one}. 
Moreover, comparing eqs.~(\ref{z-eq}) and (\ref{z'_dec}), which have different 
dependences on $x$, we find that, for $x$ smaller than a typical value 
$x_{\rm eq}$ expressed by
\be
x_{\rm eq} \approx 0.046(\Omega_m h^2)^{-1} \;, 
\ee  
the mirror photons would decouple yet during the radiation dominated period 
(see fig.~\ref{figzz}). 
Assuming, e.g., the value $\Omega_m h^2 = 0.135$ (the result of a recent 
WMAP fit \cite{wmap-par}), we obtain $x_{\rm eq} \approx 0.34$, which 
indicates that below about this value the mirror decoupling happens in the 
radiation dominated period, with consequences on structure formation 
(as we will see in the following sections and in paper 2 \cite{paper2}).

We have shown that mirror baryons could provide a significant contribution 
to the energy density of the Universe and thus they could constitute a 
relevant component of dark matter. 
Immediate question arises: how the mirror baryonic dark matter (MBDM) 
behaves and what are the differences from the more familiar dark matter 
candidates as the cold dark matter (CDM), the hot dark matter (HDM), etc.
In next sections we discuss the problem of the cosmological structure 
formation in the presence of M baryons as a dark matter component.


\section{\bf  Mirror baryonic structure formation }
\label{baryo_struct_form}

In this section, we extend the linear structure formation theory (for the 
standard scenario, see refs.~\cite{padmbook} and \cite{tsagas}) to the 
case of dark matter with a non-negligible mirror baryonic component.

In a Mirror Universe we assume that a mirror sector is present, so that the 
matter is made of ordinary baryons (the only certain component), 
non-baryonic (dark) matter, and mirror baryons. 
Thus, it is necessary to study the structure formation in all these three 
components. 
We proceed in this way: first of all, we recall the situation for ordinary 
baryons, and then we compute the same quantities for mirror baryons, 
comparing them with each other and with the CDM case.

In general, when dealing with the pre-recombination plasma, we distinguish 
between two types of perturbations, namely between ``isoentropic" 
({\sl adiabatic}) and ``entropic" ({\sl isocurvature} or {\sl isothermal}) modes 
\cite{Z}, while after matter-radiation decoupling perturbations evolve in the 
same way regardless of their original nature. 

In this paper we study only adiabatic perturbations, which are today the 
preferred perturbation modes, and we leave out the isocurvature modes, 
which could also have a contribution, but certainly cannot be the dominant 
component \cite{isocurv}. 
Here we remember only that an adiabatic perturbation satisfies the 
{\sl condition for adiabaticity}
\begin{equation}
\delta_m = {3\over4} \delta_r \;, 
\label{adcond}
\end{equation}
which, defining, as usual, $\delta \equiv \left( {\delta\rho} / \rho \right)$, 
relates perturbations in matter ($ \delta_m $) and radiation ($ \delta_r $) 
components.

We will now consider cosmological models where 
baryons, ordinary or mirror, are the dominant form of matter.
Thus, it is crucial to study the interaction between baryonic matter and 
radiation during the plasma epoch in both sectors, and the simplest way of 
doing it is by looking at models containing only these two matter 
components.

According to the Jeans theory \cite{jeans}, the relevant scale 
for the gravitational instabilities is characterized by the Jeans scale (length 
and mass), which now needs to be defined in both the ordinary and mirror 
sectors. 
Then, we define the ordinary and mirror Jeans lengths as
\begin{equation}
\lambda_{\rm J} \simeq v_{\rm s} \sqrt{ \pi \over { G\rho_{\rm dom} } } ~~~~~~~~~~~~~~
\lambda'_{\rm J} \simeq v'_{\rm s} \sqrt{ \pi \over { G\rho_{\rm dom} } } \;, 
\label{Jl1}
\end{equation}
where $\rho_{\rm dom}$ is the density of the dominant species, and 
$ v_{\rm s} $, $ v'_{\rm s} $ are the sound speeds, and the Jeans masses 
as
\begin{equation}
M_{\rm J} = {4\over3}\pi\rho_{\rm b}\left({\lambda_{\rm J}} \over {2}\right)^3 
 = {\pi\over6}\rho_{\rm b}\left({\lambda_{\rm J}}\right)^3 ~~~~~~~~~~~~~~
M'_{\rm J} = {4\over3}\pi\rho'_{\rm b}\left({\lambda'_{\rm J}} \over {2}\right)^3 \;,
\label{Jm1}
\end{equation}
where the density is now that of the perturbed component (ordinary or mirror 
baryons).


\subsection{Evolution of the adiabatic sound speed}
\label{evol_Vs}

Looking at the expressions of the Jeans length (\ref{Jl1}), it is clear that the 
key issues are the evolutions of the sound speeds in both sectors, since they determine 
the scales of gravitational instabilities. 
Using the definition of adiabatic sound speed, we obtain for the two sectors
\be \label{assom}
  v_{\rm s} = {\left({\partial p} \over {\partial \rho} \right)}^{1/2} = w^{1/2} ~~~~~~~~~~~~~~
  v_{\rm s}'= {\left({\partial p'} \over {\partial \rho'} \right)}^{1/2} = (w')^{1/2} \;,
\ee
where $ w $ and $ w' $ are relative respectively to the ordinary and mirror 
equations of state $ p = w \rho $ and $ p' = w' \rho' $.

First of all, we consider the standard case of a Universe made of only one 
sector. 
In a mixture of radiation and baryonic matter the total density and pressure 
are $ \rho = \rho_\gamma + \rho_b $ and 
$ p \simeq p_\gamma = \rho_\gamma /3 $ respectively (recall that 
$ p_{\rm b} \simeq 0 $). 
Hence, the adiabatic sound speed is given by
\begin{equation}
v_{\rm s} = \left({{\partial p} \over {\partial \rho}}\right)^{1/2}
  \simeq {1\over\sqrt{3}}\left(1+{{3\rho_b} \over 
  {4\rho_\gamma}}\right)^{-{1/2}} \;, \label{vsord}
\end{equation}
where we have used the adiabatic condition (\ref{adcond}).
In particular, using the scaling laws $ \rho_m \propto \rho_{0m}(1+z)^3 $ and 
$ \rho_\gamma \propto \rho_{0\gamma}(1+z)^4 $, together with the definition 
of matter-radiation equality (where we now consider only baryons and 
photons), we obtain
\be \label{ass2} 
v_{\rm s}(z) \simeq {1 \over {\sqrt3}}
 \left[ 1 +{3 \over 4} \left({{1+z_{\rm eq}} \over {1+z}}\right)\right]^{-1/2} \;. 
\ee 
In fact, the relation above is valid only for an ordinary Universe, and it is an 
approximation, for small values of $x$ and the mirror baryon density 
(remember that $\beta = {\Omega_b' / \Omega _b}$), of the more general 
equation for a Universe made of two sectors of baryons and photons, 
obtained using eqs.~(\ref{vsord}) and (\ref{z-eq_2}) and given by
\be \label{ass3} 
v_{\rm s}(z) \simeq {1 \over {\sqrt3}}
\left[ 1 +{3 \over 4}  \left({1+x^{4}} \over {1+\beta } \right) 
\left( {{1+z_{\rm eq}} \over {1+z}} \right) \right]^{-1/2} \;.
\ee 
In the most general case, the matter is made not only of ordinary and mirror 
baryons, but also of some other form of dark matter, so the factor $1+\beta $ 
is replaced by $1+\beta +\beta _{DM}$, where 
$\beta _{DM} 
= {\left(\Omega _m - \Omega _b - \Omega _b' \right) / \Omega _b}$. 
The presence of the term $ [(1 + x^4) / (1 + \beta) ] $ in equation (\ref {ass3}) 
is linked to the shift of matter-radiation equality epoch (\ref {shifteq}), thus it 
only balance this effect, without changing the value of the sound speed 
computed using eq. (\ref {ass2}).

The mirror plasma contains more baryons and less photons than the 
ordinary one, $ \rho'_b = \beta \rho_b $ and 
$ \rho'_\gamma = x^4 \rho_\gamma $.  
Then, using eqs.~(\ref {assom}) and (\ref {z-eq_2}), we have
\be \label{mirsound} 
v'_{\rm s}(z) \simeq 
  {1 \over {\sqrt3}} \left(1+ {{3\rho'_b} \over {4\rho'_\gamma}}\right)^{-1/2} 
  \approx {1 \over {\sqrt3}} \left[ 1 +{3 \over 4} \left({{1+x^{-4}} \over 
  {1+\beta ^{-1}}}\right) \left({{1+z_{\rm eq}} \over {1+z}}\right)\right]^{-1/2} \;. 
\ee 
Let us consider for simplicity the case when dark matter of the Universe is 
entirely due to M baryons, $ \Omega_m \simeq \Omega'_b $ 
(i.e., $ \beta \gg 1 $). 
Hence, for the redshifts of cosmological relevance, $ z \sim z_{\rm eq} $, 
we have $ v'_{\rm s} \sim 2x^2 /3 $, which is always less than 
$ v_{\rm s} \sim 1/\sqrt{3} $ 
(some example: if $ x = 0.7 $, $ v_{\rm s}' \approx 0.5 \cdot v_{\rm s} $; 
if $ x = 0.3 $, $ v_{\rm s}' \approx 0.1 \cdot v_{\rm s} $). 
In expression (\ref{mirsound}) it is crucial the presence of the factor 
$[(1+x^{-4}) / (1+\beta ^{-1})]$, which is always greater than 1 
(given the bounds of eqs.~(\ref{mir-par}) on the parameters), so that  
$ v'_{\rm s} < v_{\rm s} $ during all the history of the Universe, and only 
in the limit of very low scale factors, $ a \ll a_{\rm eq} $, we obtain 
$ v'_{\rm s} \simeq v_{\rm s} \simeq 1/\sqrt{3} $. 
As we will see in the following, this has important consequences on 
structure formation scales.

\begin{figure}[h]
  \begin{center}
    \leavevmode
    \epsfxsize = 10cm
    \epsffile{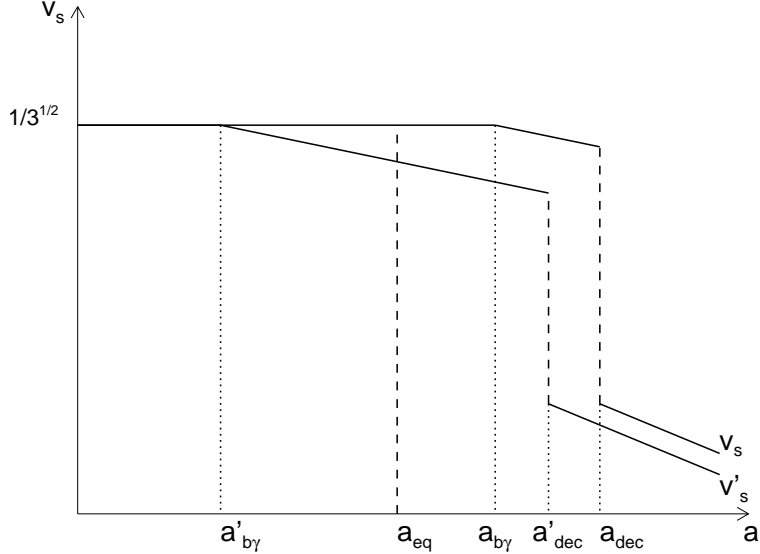}
  \end{center}
\caption{\small The trends of the mirror sound speed ($v_{\rm s}'$) as a 
function 
of the scale factor, compared with the ordinary sound speed ($v_{\rm s}$). 
The ordinary model has $ \Omega_b h^2 = 0.08 $, 
while the mirror model has $ x = 0.6 $ and $ \beta = 2 $. 
Are also reported all the key epochs: photon-baryon equipartition and 
decoupling in both sectors, and the matter-radiation equality. }
\label{scalord2}
\end{figure}

Now we define $ a_{\rm b\gamma } $ as the scale factor corresponding to 
the redshift 
\be
(1+z_{\rm b\gamma }) = (a_{\rm b\gamma})^{-1} = 
  (\Omega _b / \Omega _\gamma ) = 3.9\cdot 10^4 (\Omega _b h^2) \;. 
\ee
Since $ 1+z_{\rm rec} \simeq 1100 $, ordinary baryon-photon equipartition 
occurs before recombination only if $ \Omega_b h^2>0.026 $ (which seems 
unlikely, given its current estimates). 
According to what found in \S~\ref{mir_dm}, in the mirror sector the scale 
of baryon-photon equality $ a_{\rm b\gamma}' $ is dependent on $ x $ and 
it transforms as
\be \label{shiftabg} 
a_{\rm b\gamma}' 
  = { \Omega_{\gamma}' \over \Omega_b' } 
  \simeq { \Omega_{\gamma} \, x^4 \over \Omega_b \, \beta} 
  = a_{\rm b\gamma} { x^4 \over \beta } < a_{\rm b\gamma} \;.
\ee
If we remember the definition of the quantity 
$ x_{\rm eq} \approx 0.046 (\Omega_m h^2)^{-1} $, we have that for 
$ x > x_{\rm eq} $ the decoupling occurs after equipartition (as in the 
ordinary sector for $ \Omega_b h^2>0.026 $), while for $ x < x_{\rm eq} $ 
it occurs before (as for $ \Omega_b h^2 < 0.026 $). 

Regardless of which sector we are considering, in the radiation era 
$ \rho_\gamma \gg \rho_b $, ensuring that $ v_{\rm s} \simeq 1/\sqrt{3} $. 
In the interval between equipartition and decoupling, when 
$ \rho_b \gg \rho_\gamma $, eq.~(\ref{vsord}) gives 
$ v_{\rm s} \simeq \sqrt{4 \rho_\gamma / 3 \rho_b}\propto a^{-1/2} $. 
After decoupling there is no more pressure equilibrium between baryons 
and photons, and $ v_{\rm s} $ is just the velocity dispersion of a gas of 
hydrogen and helium, $ v_{\rm s} \propto a^{-1} $. 
If $ \Omega_b h^2 < 0.026 $ or $ x < x_{\rm eq} $ (according to what 
sector we consider), photon-baryon equipartition occurs after decoupling, 
and the intermediate situation does not arise. 

\begin{figure}[h]
  \begin{center}
    \leavevmode
    \epsfxsize = 10cm
    \epsffile{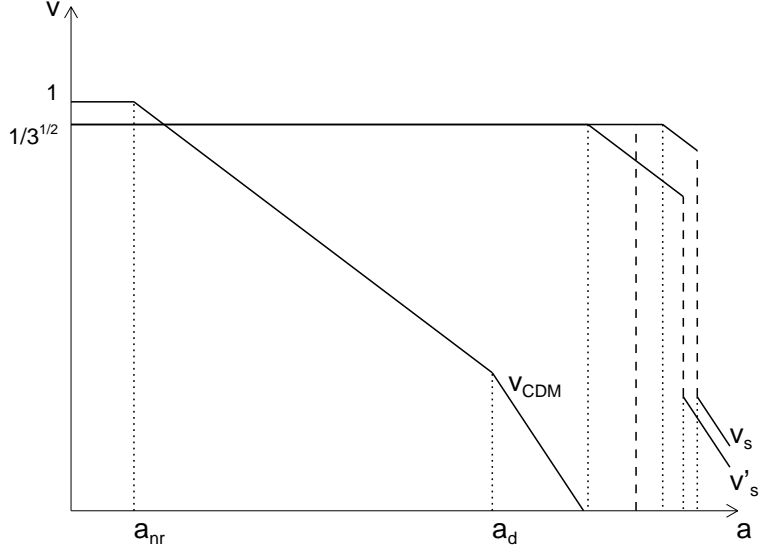}
  \end{center}
\caption{\small The trends of the mirror ($ v_{\rm s}' $) and ordinary 
($ v_{\rm s} $) sound speed compared with the velocity dispersion of a 
typical non baryonic cold dark matter candidate of mass 
$ \sim 1 \, {\rm GeV} $ ($ v_{\rm CDM} $); $ a_{\rm nr} $ and $ a_{\rm d} $ 
indicate the scale factors at which the dark matter particles become 
non relativistic or decouple. 
The ordinary and mirror models are the same as in figure \ref{scalord2}, 
but the horizontal scale is expanded by some decade in order to show 
the CDM velocity.}
\label{scalord3}
\end{figure}

It follows that, by taking care to interchange $ a_{\rm b\gamma} $ with 
$ a_{\rm b\gamma}' $ and $ a_{\rm dec} $ with $ a_{\rm dec}' $, we have 
for the sound speed the same trends with the scale factor in both sectors, 
though with the aforementioned differences in the values. 
The situation whit $ x > x_{\rm eq} $ is resumed below:
\begin{equation}
v'_{\rm s}(a) \propto\left\{\begin{array}{l}
 const. \hspace{11mm} a<a'_{b\gamma} \;,\\
 a^{-1/2} \hspace{12mm} a'_{b\gamma}<a<a'_{\rm dec} \;,\\
 a^{-1}  \hspace{15mm} a>a'_{dec} \;.\\\end{array}\right.  \label{vsz}
\end{equation}

If we recall that the matter-radiation equality for a single sector (ordinary) 
Universe, 
$ (a_{\rm eq})_{\rm ord} $, is always bigger than that for a two sectors 
(mirror) one, $ (a_{\rm eq})_{\rm mir} $, according to
\be \label{shiftaeq} 
(a_{\rm eq})_{\rm mir} 
  = { {\left( 1+x^4 \right)} \over {\left( 1+\beta \right)} } ~ (a_{\rm eq})_{\rm ord} 
  < (a_{\rm eq})_{\rm ord} \;,
\ee
together with our hypothesis $ x < 1 $ (from the BBN bound) 
and $ \beta > 1 $ (cosmologically interesting situation, i.e., significant 
mirror baryonic contribution to the dark matter), we obtain the useful 
inequality always verified in a Universe made of O and M sectors 
\be \label{hiera1} 
a_{\rm b\gamma}' < a_{\rm eq} < a_{\rm b\gamma} \;.
\ee

It's very important to remark (see ref.~\cite{padmbook}) 
that at decoupling $ v_{\rm s}^2 $ drops from $ (p_\gamma / \rho _b) $ to 
$ (p_b /\rho _b) $. 
Since $ p_\gamma \propto n_\gamma T $ while $ p_b \propto n_b T $ with 
$ (n_\gamma /n_b) \simeq 10^9 \gg 1 $ in the ordinary sector, this is a 
large drop in $ v_{\rm s} $ and consequently in $ \lambda _{\rm J} $. 
More precisely, $ v_{\rm s}^2 $ drops from the value 
$ (1/3)(\rho _\gamma / \rho _b) 
= (1/3)(\Omega _\gamma / \Omega _b)(1+z_{dec}) $ to the value 
$ (5/3) (T_{dec}/m_b) = (5/3) (T_0/m_b)(1+z_{dec}) $, with a reduction factor 
\be
F_1 (\Omega_b h^2 > 0.026) 
= { (v_{\rm s}^2)^{\rm (+)}_{\rm dec} \over (v_{\rm s}^2)^{\rm (-)}_{\rm dec} } 
= 6.63 \cdot 10^{-8} (\Omega _b h^2) \;,
\ee 
where $ (v_{\rm s}^2)^{\rm (-)}_{\rm dec} $ and 
$ (v_{\rm s}^2)^{\rm (+)}_{\rm dec} $ indicate the sound speed respectively 
just before and after decoupling.
If we consider now the mirror sector and the drop in $ (v_{\rm s}')^2 $ at 
decoupling, we find for the reduction factor
\be \label{f1p}
F_1' (x > x_{\rm eq}) = \beta x^{-3} F_1 \;.
\ee
In the case $ \Omega_b h^2 < 0.026 $, $ v_{\rm s}^2 $ drops directly from 
$ (1/3) $ to $ (5/3) (T_{dec}/m_b) = (5/3) (T_0/m_b)(1+z_{dec}) $ with a 
suppression
\be
F_2 (\Omega_bh^2 < 0.026) = 
  { (v_{\rm s}^2)^{\rm (+)}_{\rm dec} \over (v_{\rm s}^2)^{\rm (-)}_{\rm dec} } 
  = 1.9 \cdot 10^{-9} \;.
\ee
It's easy to find that in the mirror sector the reduction factor in the case 
$ a_{b\gamma}' > a_{\rm dec}' $ is the same as in the ordinary one
\be \label{f2p}
F_2' (x < x_{\rm eq}) = F_2 \;.
\ee
Some example: for $ x = 0.7 $, $ F_1' \approx 2.9 \beta F_1 $; 
for $ x = 0.5 $, $ F_1' = 8 \beta F_1 $; 
for $ x = 0.3 $, $ F_1' \approx 37 \beta F_1 $. 
We remark that, if $ \beta \geq 1 $, $ F_1' $ is at least about an order of 
magnitude larger than $ F_1 $. 
In fact, after decoupling 
$ (v_{\rm s}')^2 = (5/3) (T_{\rm dec}' / m_b) = (5/3) (T_{\rm dec} / m_b) 
= (v_{\rm s})^2$ (since $T_{\rm dec}' = T_{\rm dec} $), and between 
equipartition and recombination $ (v_{\rm s}')^2 < (v_{\rm s})^2 $. 
The relation above means that the drop is smaller in the mirror sector than 
in the ordinary one.
Obviously, before equipartition $ (v_{\rm s}')^2 = (v_{\rm s})^2 = 1/3 $, and 
for this reason the parameter $ F_2 $ is the same in both sectors.

In figure \ref{scalord2} we plot the trends with scale factor of the mirror 
sound speed, in comparison with the ordinary one. 
The ordinary model is a typical one with $ \Omega_b h^2 > 0.026 $, while 
the mirror model has $ x = 0.6 $ and $ \beta = 2 $ (this means that M 
baryonic density is twice the ordinary one, chosen in these models about 
four times its current estimation in order to better show the general 
behaviour). 
In the same figure we show also the aforementioned relative positions 
of the key epochs (photon-baryon equipartition and decoupling) for both 
sectors, together with the matter-radiation equality.
If we reduce the value $ \Omega_b h^2 $, $ a_{\rm b\gamma} $ goes 
toward higher values, while $ a_{\rm dec} $ remains fixed, so that for 
$ \Omega_b h^2 < 0.026 $ decoupling happens before equipartition and 
the intermediate regime, where $ v_{\rm s} \propto a^{-1/2} $, disappears.
In an analogous way, if we reduce $ x $, $ a'_{\rm dec} $ shifts to lower 
values, until for $ x < x_{\rm eq} $ it occurs before the mirror equipartition
$ a'_{\rm b\gamma} $, so that the intermediate regime for $ v'_{\rm s} $ 
disappears.

In figure \ref{scalord3} the same ordinary and mirror sound speeds are 
plotted together with the velocity dispersion of a typical non baryonic cold 
dark matter candidate of mass $ \sim $ 1GeV. 
Note that the horizontal scale is expanded by some decade compared to 
the figure \ref{scalord2}, because the key epochs for the CDM velocity 
evolution (the epochs when the particles become non relativistic, 
$ a_{\rm nr} $, and when they decouple, $ a_{\rm d} $) occur at much 
lower scale factors.


\begin{figure}[h]
  \begin{center}
    \leavevmode
    \epsfxsize = 10cm
    \epsffile{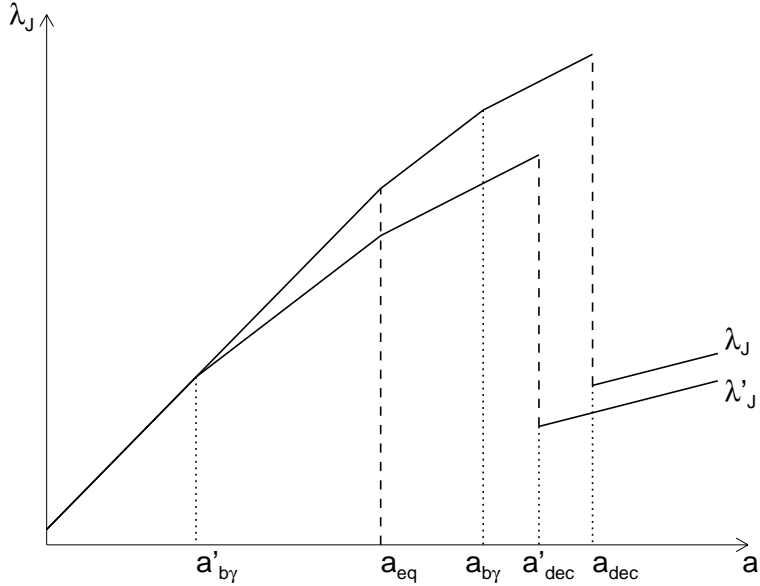}
  \end{center}
\caption{\small The trends of the mirror Jeans length ($ \lambda'_{\rm J} $) 
as a function of the scale factor, compared with the ordinary Jeans length 
($ \lambda_{\rm J} $). 
The ordinary model has $ \Omega_b h^2 = 0.08 $, while the mirror model 
has $ x = 0.6 $ and $ \beta = 2 $. 
The horizontal scale is the same as in figure \ref{scalord2}.
We remark that the same behaviours of the ordinary sector are present 
in the mirror sector for different intervals of scale factor.}
\label{scalord6}
\end{figure}

\subsection{Evolution of the Jeans length and the Jeans mass}

Recalling the definitions (\ref{Jl1}) and (\ref{Jm1}), 
and using the results of \S~\ref{evol_Vs} relative to the sound speed, 
we can compute the evolution of the mirror Jeans length and mass and 
compare them with the analogous quantities for ordinary baryons and CDM.

We find for the evolution of the adiabatic Jeans length and mass of ordinary 
baryons in the case $ \Omega_b h^2 > 0.026 $
\begin{equation}
\lambda_{\rm J}\propto\left\{\begin{array}{l}
 a^2\,\\ 
 a^{3/2}\,\\ 
 a\,\\
 a^{1/2} \,
 \\\end{array}\right.  \label{aJl}
\hspace{12mm}M_{\rm J}\propto{{\lambda _{\rm J}^3} \over {a^3}}
\propto\left\{\begin{array}{l}
 a^3\hspace{20mm}a<a_{\rm eq}\,\\ 
 a^{3/2}\hspace{17mm}a_{\rm eq}<a<a_{\rm b\gamma }\,\\ 
 const.\hspace{13mm}a_{\rm b\gamma }<a<a_{\rm dec}\,\\
 a^{-3/2}\hspace{15mm}a_{\rm dec}<a \,
\end{array}\right.  \label{aJlm}
\end{equation}

\noindent Otherwise, if $ \Omega_b h^2 \leq 0.026 $, 
$a_{\rm b\gamma } > a_{\rm dec}$, there is no intermediate phase 
$a_{\rm b\gamma } < a < a_{\rm dec}$. 

In the mirror sector it's no more sufficient to interchange 
$ a_{\rm b\gamma } $ with $ a_{\rm b\gamma }' $ and $ a_{\rm dec} $ with 
$ a_{\rm dec}' $, as made for the sound speed, because from relation 
(\ref{hiera1}) we know that in the mirror sector the photon-baryon 
equipartition happens before the matter-radiation equality (due to the fact 
that we are considering a mirror sector with more baryons and less photons 
than the ordinary one). 
It follows that, due to the shifts of the key epochs, the intervals of scale 
factor for the various trends are different. 
As usual, there are two different possibilities, $ x > x_{\rm eq} $ and 
$ x < x_{\rm eq} $ (which correspond roughly to $ \Omega_b h^2 > 0.026 $ 
and $ \Omega_b h^2 < 0.026 $ in an ordinary Universe), where, as 
discussed in \S ~\ref{evol_Vs}, for the second one the intermediate 
situationis absent.

Using the results of \S~\ref{evol_Vs} for the sound speed, we find the 
evolution of the adiabatic Jeans length and mass in the case 
$ x > x_{\rm eq} $
\begin{equation}
\lambda_{\rm J}' \propto \left\{\begin{array}{l}
 a^2\,\\ 
 a^{3/2}\,\\ 
 a\,\\
 a^{1/2} \,
 \\\end{array}\right.
\hspace{12mm}M_{\rm J}' 
\propto {{(\lambda _{\rm J}')^3} \over {a^3}}\propto\left\{\begin{array}{l}
 a^3 \hspace{20mm} a < a_{\rm b\gamma }' \,\\ 
 a^{3/2} \hspace{17mm} a_{\rm b\gamma }' < a < a_{\rm eq} \,\\ 
 const. \hspace{13mm} a_{\rm eq} < a < a_{\rm dec}' \,\\
 a^{-3/2} \hspace{15mm} a_{\rm dec}' < a \,
\end{array}\right.
\end{equation}

\begin{figure}[h]
  \begin{center}
    \leavevmode
    \epsfxsize = 10cm
    \epsffile{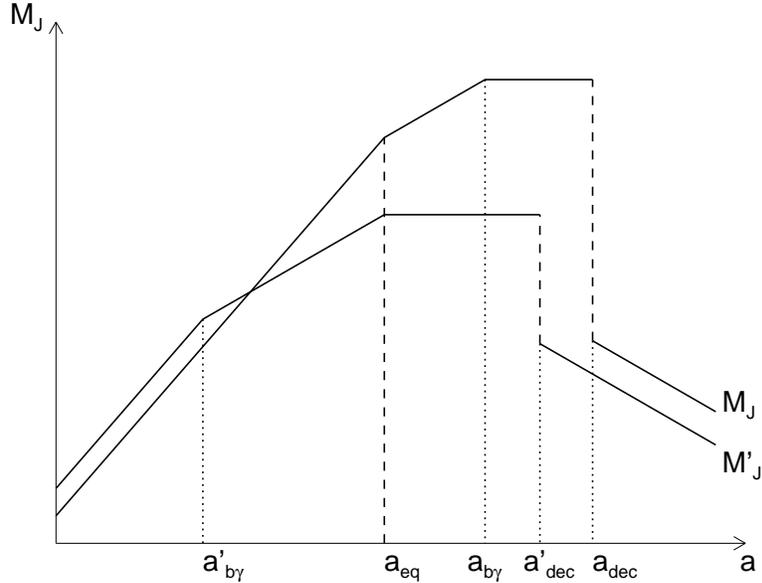}
  \end{center}
\caption{\small The trends of the mirror Jeans mass ($ M'_{\rm J} $) as a 
function of the scale factor, compared with the ordinary Jeans mass 
($ M_{\rm J} $). 
The models and the horizontal scale are the same as in figure \ref{scalord6}.}
\label{scalord8}
\end{figure}

We plot in figures \ref{scalord6} and \ref{scalord8} with the same horizontal 
scale the trends of the mirror Jeans length and mass compared with those 
for the ordinary sector; the parameters of both mirror and ordinary models 
are the ones previously used, i.e. 
$ \Omega_b h^2 = 0.08 $, $ x = 0.6 > x_{\rm eq} $ and $ \beta = 2 $. 

In the ordinary sector the greatest value of the Jeans mass is just before 
decoupling (see ref.~\cite{padmbook}), in the interval 
$ a_{\rm b\gamma } < a < a_{\rm dec} $ , where
\be
M_{\rm J}(a \lsim a_{\rm dec}) 
  = 1.47 \cdot 10^{14} M_\odot \left( 1 + \beta \right)^{-3/2} 
  \left(\Omega_{\rm b} h^2 \right)^{-2} \;,
\ee
that for an hypothetical $ \Omega_b \simeq 0.1 h^{-2} $ and $ \beta = 0 $ is 
$ \sim 10^{16} M_\odot $. 
Just after decoupling we have
\be
M_{\rm J}(a \gsim a_{\rm dec}) 
  = 2.5 \cdot 10^{3} M_\odot \left( 1 + \beta \right)^{-3/2} 
  \left(\Omega_{\rm b} h^2 \right)^{-1/2} \;,
\label{mj1after}
\ee
that for $ \Omega_b \simeq 0.1 h^{-2} $ and $ \beta = 0 $ is 
$ \sim 10^{4} M_\odot $. 
This drop is very sudden and large, changing the Jeans mass by 
$ F_1^{3/2} \simeq 1.7 \cdot 10^{-11} (\Omega _b h^2)^{3/2} $. 

\noindent Otherwise, if $\Omega_b h^2 \leq 0.026$, 
$ a_{\rm b\gamma } > a_{\rm dec} $, there is no intermediate phase 
$ a_{\rm b\gamma } < a < a_{\rm dec} $, and 
$ M_{\rm J}(a \lsim a_{\rm dec}) $ is larger
\be
M_{\rm J}(a \lsim a_{\rm dec}) 
  \simeq 3.1 \cdot 10^{16} M_\odot \left( 1 + \beta \right)^{-3/2} 
  \left(\Omega_{\rm b} h^2 \right)^{-1/2} \;,
\ee
while after decoupling it takes the value in eq.~(\ref{mj1after}), so that the 
drop is larger, $ F_2^{3/2} \simeq 8.3 \cdot 10^{-14} $.
We observe that, with the assumptions $ \Omega_b = 1 $ (totally excluded 
by observations) and $ \beta = 0 $, $ M_{\rm J,max} $ (which is the first 
scale to become gravitationally unstable and collapse soon after 
decoupling) has the size of a supercluster of galaxies.

If we now consider the expression (\ref {shiftabg}), we have
\be
{ a'_{\rm b\gamma } \over a_{\rm eq} }
  = \left( { 1 + \beta } \over { \beta } \right) 
  \left( { x^4 } \over { 1 + x^4 } \right) \;,
\ee
which can be used to express the value of the mirror Jeans mass in the 
interval $ a_{\rm eq} < a < a_{\rm dec}' $ (where $ M_{\rm J}' $ takes the 
maximum value) in terms of the ordinary Jeans mass in the corresponding 
ordinary interval $ a_{\rm b\gamma} < a < a_{\rm dec} $. 
We obtain
\be
M_{\rm J}'(a \lsim a_{\rm dec}') \approx 
  \beta^{-1/2} \left( { x^4 \over {1 + x^4} } \right)^{\rm 3/2} 
  \cdot M_{\rm J}(a \lsim a_{\rm dec}) \;,
\ee
which, for $ \beta \geq 1 $ and $x < 1$, means that the Jeans mass for the M 
baryons is lower than for the O ones over the entire permitted ($\beta$-$x$) 
parameter space, with implications for the structure formation process. 
If, e.g., $ x = 0.6 $ and $ \beta = 2 $, then 
$ M_{\rm J}' \sim 0.03 \; M_{\rm J} $. 
We can also express the same quantity in terms of $ \Omega_b $, $ x $ 
and $ \beta $, in the case that all the dark matter is in the form of mirror 
baryons, as
\be \label{mj_mir_1}
M_{\rm J}'(a \lsim a_{\rm dec}') \approx 
  3.2 \cdot  10^{14} M_\odot \;
  \beta^{-1/2} ( 1 + \beta )^{-3/2} \left( x^4 \over {1+x^4} \right)^{3/2} 
  ( \Omega_{\rm b} h^2 )^{-2} \;.
\ee

If we remember eq.~(\ref {f1p}), we obtain that for the mirror model the drop 
in the Jeans 
mass at decoupling is $ (F_1')^{3/2} = \beta^{3/2} x^{-9/2} (F_1)^{3/2} $, which, 
given our bounds on $ x $ and $ \beta $, is greater than $ (F_1)^{3/2} $. 
We give here some numerical example: for $ x = 0.7 $, 
$ (F_1')^{3/2} \approx 5 \beta^{3/2} (F_1)^{3/2} $; 
for $ x = 0.6 $ and $ \beta = 2 $ (the case of figures \ref{scalord6} and 
\ref{scalord8}), $ (F_1')^{3/2} \approx 28 (F_1)^{3/2} $.

It's important to stress that these quantities are strongly 
dependent on the values of the free parameters $ x $ and $ \beta $, which 
shift the key epochs and change their relative positions. 
We can describe some case useful to understand the general behaviour, but 
if we want an accurate solution of a particular model, we must 
unambiguously identify the different regimes and solve in detail the 
appropriate equations.

\begin{figure}[h]
  \begin{center}
    \leavevmode
    \epsfxsize = 10cm
    \epsffile{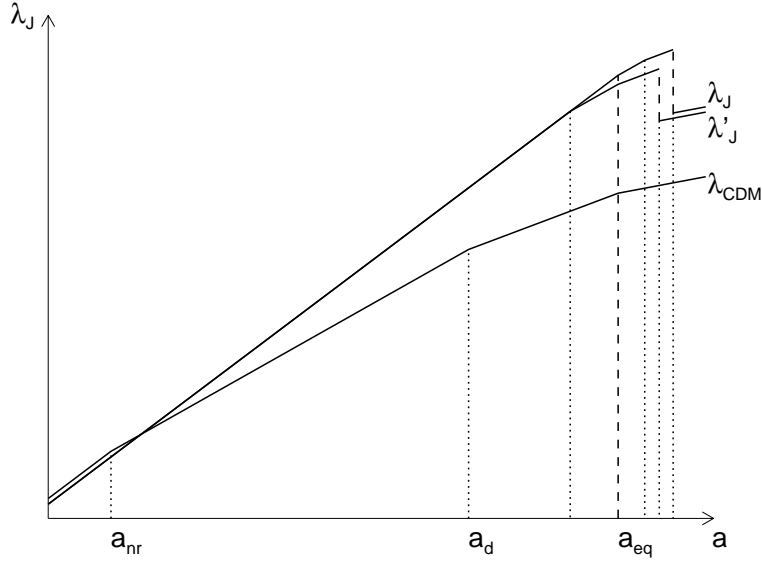}
  \end{center}
\caption{\small The trends of the mirror ($ \lambda'_{\rm J} $) and ordinary 
($ \lambda_{\rm J} $) Jeans 
length compared with that of a typical non baryonic cold dark matter 
candidate of mass $ \sim 1 \, {\rm GeV} $ ($ \lambda'_{\rm CDM} $); 
$ a_{\rm nr} $ and $ a_{\rm d} $ indicate the scale factors at which the dark 
matter particles become non relativistic or decouple, respectively. 
The models are the same as in figure \ref{scalord6}, but the horizontal scale 
is expanded by some decade to show the CDM Jeans length, as in figure 
\ref{scalord3}.}
\label{scalord7}
\end{figure}

\begin{figure}[h]
  \begin{center}
    \leavevmode
    \epsfxsize = 10cm
    \epsffile{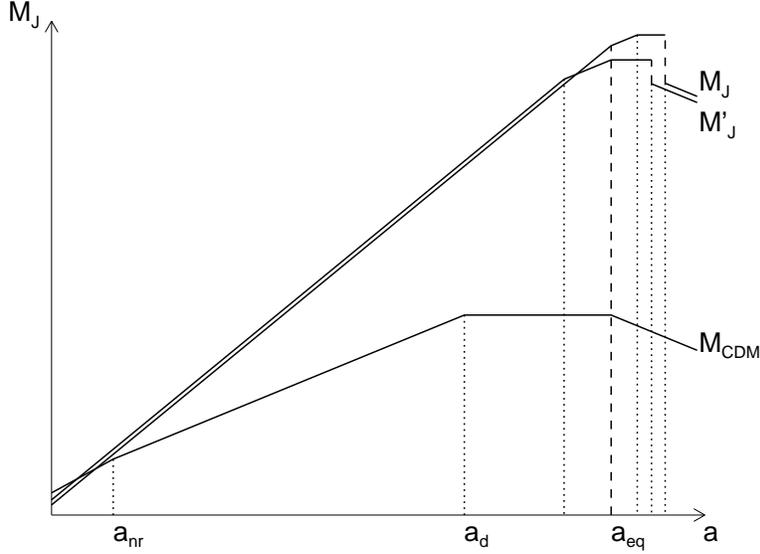}
  \end{center}
\caption{\small The trends of the mirror ($ M'_{\rm J} $) and ordinary 
($ M_{\rm J} $) Jeans mass compared with those of a typical non baryonic 
cold dark matter candidate of mass $ \sim 1 \, {\rm GeV} $ ($ M_{\rm CDM} $); 
$ a_{\rm nr} $ and $ a_{\rm d} $ indicate the scale factors at which the dark 
matter particles become non relativistic or decouple, respectively. 
The model parameters and the horizontal scale are the same as in figure 
\ref{scalord7}.}
\label{scalord9}
\end{figure}

In figures \ref{scalord7} and \ref{scalord9} we plot the trends of the mirror 
and ordinary Jeans length and mass compared with those of a typical non 
baryonic cold dark matter candidate of mass $ \sim $ 1GeV. 
Apart from the usual expansion of the horizontal scale, due to the much 
lower values of the CDM key epochs as compared to the baryonic ones, a 
comparison of the mirror scenario with the cold dark matter one shows that 
the maximal value of the CDM Jeans mass is several orders of magnitude 
lower than that for mirror baryons.
This implies that a very large range of mass scales, which in a mirror 
baryonic scenario oscillate before decoupling, in a cold dark matter scenario 
would grow unperturbed during all the time (for more details see 
\S~\ref{mirevolpert}).

For the case $ x < x_{\rm eq} $, both $ a_{\rm b\gamma}' $ and 
$ a_{\rm dec}' $ are smaller than the previous case $ x > x_{\rm eq} $, while 
the matter-radiation equality remains practically the same; as explained in 
\S~\ref{mir_dm}, the mirror decoupling (with the related drop in the 
associated quantities) happens before the matter-radiation equality, and the 
trends of the mirror Jeans length and mass are the following
\begin{equation}
\lambda_{\rm J}' \propto \left\{\begin{array}{l}
 a^2\,\\ 
 a^{3/2}\,\\ 
 a\,\\
 a^{1/2} \,
 \\\end{array}\right.
\hspace{12mm}M_{\rm J}' \propto {{(\lambda _{\rm J}')^3} \over {a^3}}
\propto\left\{\begin{array}{l}
 a^3 \hspace{20mm} a < a_{\rm b\gamma }' \,\\ 
 a^{3/2} \hspace{17mm} a_{\rm b\gamma }' < a < a_{\rm dec}' \,\\ 
 const. \hspace{13mm} a_{\rm dec}' < a < a_{\rm eq} \,\\
 a^{-3/2} \hspace{15mm} a_{\rm eq} < a \,
\end{array}\right.
\end{equation}

\begin{figure}[h]
  \begin{center}
    \leavevmode
    \epsfxsize = 10cm
    \epsffile{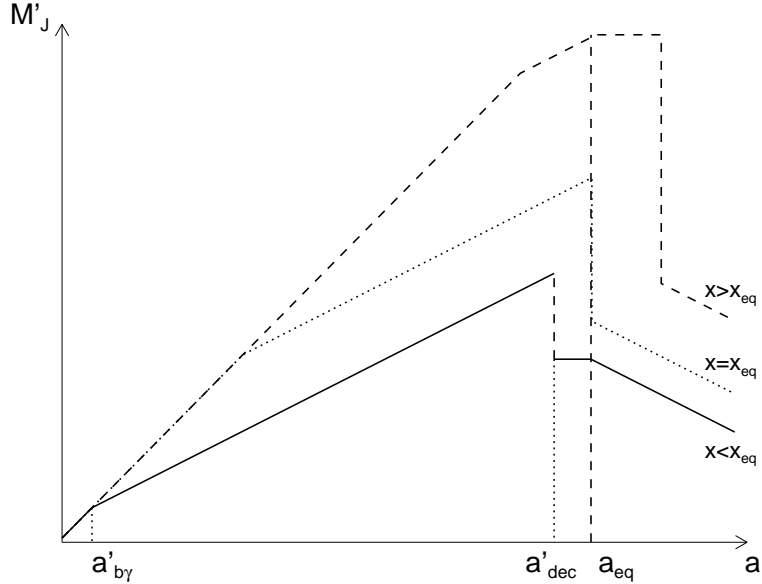}
  \end{center}
\caption{\small The trends of the mirror Jeans mass for the cases 
$ x < x_{\rm eq} $ (solid line), $ x = x_{\rm eq} $ (dotted) and 
$ x > x_{\rm eq} $ (dashed). The model with $ x > x_{\rm eq} $ is the same 
as in figure \ref{scalord8}, the others are obtained changing only the value 
of $ x $ and keeping constant all the other parameters. 
As clearly shown in the figure, the only key epoch which remains almost 
constant in the three models is the matter-radiation equality; the mirror 
baryon-photon equipartition and decoupling indicated are relative to the 
model with $ x < x_{\rm eq} $. 
It's also evident the change in the trends when $ x $ becomes lower than 
$ x_{\rm eq} $, due to the fact that $ a_{\rm dec}' $ becomes lower than 
$ a_{\rm eq} $.}
\label{scalord10}
\end{figure}

In this case we obtain for the highest value of the Jeans mass just 
before decoupling the expression
\be \label{mj_mir_2}
M_{\rm J}'(a \lsim a_{\rm dec}') \approx 
  3.2 \cdot  10^{14} M_\odot \; 
  \beta^{-1/2} ( 1 + \beta )^{-3/2} 
  \left( x \over x_{\rm eq} \right)^{3/2} \left( x^4 \over {1+x^4} \right)^{3/2} 
  ( \Omega_{\rm b} h^2 )^{-2} \;.
\ee
In case $ x = x_{\rm eq} $, the expressions 
(\ref{mj_mir_1}) and (\ref{mj_mir_2}), respectively valid for 
$ x \ge x_{\rm eq} $ and $ x \le x_{\rm eq} $, are coincident, as we expect.
If we consider the differences between the highest mirror Jeans mass for 
the particular values $ x = x_{\rm eq}/2 $, $ x = x_{\rm eq} $ 
and $ x = 2 x_{\rm eq} $, we obtain the following relations:
\be
M_{\rm J,max}'(x_{\rm eq}/2) \approx \left( 1 \over 2 \right)^{15/2} 
  \left( {1 + x_{\rm eq}^4} 
  \over {1 + \left(x_{\rm eq} / 2 \right)^4} \right)^{3/2} 
  M_{\rm J,max}'(x_{\rm eq}) 
  \approx 0.005 \: M_{\rm J,max}'(x_{\rm eq}) \;,
\ee
\be
M_{\rm J,max}'(2x_{\rm eq}) \approx 2^6 
  \left( {1 + x_{\rm eq}^4} 
  \over {1 + (2 x_{\rm eq})^4} \right)^{3/2} 
  M_{\rm J,max}'(x_{\rm eq}) 
  \approx 64 \: M_{\rm J,max}'(x_{\rm eq}) \;.
\ee
In figure \ref{scalord10} we plot the mirror Jeans mass for the three different 
possibilities: $ x < x_{\rm eq} $, $ x > x_{\rm eq} $ and $ x = x_{\rm eq} $ 
(the transition between the two regimes), 
keeping constant all other parameters.
In these three models the matter-radiation equality is the only key epoch 
which remains almost constant. 
The change in the trends when $ x $ becomes lower than 
$ x_{\rm eq} $, due to the fact that $ a_{\rm dec}' $ becomes lower than 
$ a_{\rm eq} $, generates an evident decrease of the Jeans mass.

\subsection{Evolution of the Hubble mass}
\label{subsec_evol_MH2}

The trends of the Hubble length and mass are expressed, as usually, by
\begin{equation}
\lambda_{\rm H} \propto \left\{\begin{array}{l}
 a^2\,\\ 
 a^{3/2}\,
 \\\end{array}\right.
\hspace{12mm}M_{\rm H} \propto {{(\lambda_{\rm H})^3} \over {a^3}}
\propto\left\{\begin{array}{l}
 a^3 \hspace{20mm} a < a_{\rm eq} \,\\ 
 a^{3/2} \hspace{17mm} a > a_{\rm eq} \,
\end{array}\right.
\end{equation}

It should be emphasized that, as for the ordinary baryons, 
during the period of domination of photons ($ a < a'_{\rm b\gamma } $) the 
mirror baryonic Jeans mass is of the same order of the Hubble mass. 
In fact, following definitions we find
\be
{{M'_{\rm J}}\over{M_{\rm H}}} 
= {({\lambda'_{\rm J})^3}\over{\lambda_{\rm H}^3}} 
\simeq 26 \;.
\ee

We plot the trends of the Hubble mass in figures \ref {mirmasssca1} and 
\ref{mirmasssca2} together with other fundamental mass scales.


\subsection{Dissipative effects: collisional damping}
\label{disseff-colldamp}

A peculiar feature of the mirror baryonic scenario is that mirror baryons
undergo the collisional damping as ordinary ones.
This dissipative process modify the purely gravitational evolution of 
perturbations. 
The physical phenomenon is the interaction between baryons and photons 
before the recombination, and the consequent dissipation due to viscosity 
and heat conduction.
Around the time of recombination the perfect fluid approximation breaks 
down, and the perturbations in the photon-baryon plasma suffer from 
collisional damping. 
As decoupling is approached, the photon mean free path increases and 
photons can diffuse from the overdense into the underdense regions, 
thereby smoothing out any inhomogeneities in the photon-baryon plasma. 
This effect is known as Silk damping \cite{silk}. 

In order to obtain an estimate of the effect, we follow ref.~\cite{kolbbook} for 
ordinary baryons, and then we extend to mirror baryons.
We consider the photon mean free path
\begin{equation}
\lambda _{\gamma} = {{1} \over {X_{\rm e}{\rm n}_{\rm e}\sigma_{\rm T}}} 
\simeq 10^{29}a^3X_{\rm e}^{-1}\left(\Omega_{\rm b}h^2\right)^{-1}\,{\rm cm} \;,  \label{pmp}
\end{equation}
where $ X_{\rm e} $ is the electron ionization factor, 
$ {\rm n}_{\rm e} \propto a^{-3} $ is the number density of the free electrons 
and $ \sigma_{\rm T} $ is the cross section for Thomson scattering. 
Clearly, photon free streaming should completely damp all baryonic 
perturbations with wavelengths smaller than $ \lambda_\gamma $. 
Damping, however, occurs on scales much larger than 
$ \lambda_\gamma $ since the photons slowly diffuse from the overdense 
into the underdense regions, dragging along the still tightly coupled baryons. 
Integrating up to decoupling time we obtain the total distance 
traveled by a typical photon
\begin{equation}
\lambda _{\rm S}= 
  \sqrt{ {3\over5} {{ (\lambda _\gamma)_{\rm dec} t_{\rm dec} }
  \over a^2_{\rm dec} } } \,\,
  \simeq\,\,3.5\left(\Omega_b h^2\right)^{-3/4}\,{\rm Mpc} \;,
\label{lS}
\end{equation}
and the associated mass scale, the Silk mass, given by
\begin{equation}
M_{\rm S} 
  = {4\over3}\pi\rho_b\left({{\lambda _{\rm S}} 
  \over {2}}\right)^3\simeq6.2\times10^{12}
  \left(\Omega_b h^2\right)^{-5/4}\,{\rm M}_{\odot} \;,\label{Sm}
\end{equation}
which, assuming $ \Omega_b h^2 \simeq 0.02 $, 
gives $ M_{\rm S} \simeq 8 \times10^{14}~{\rm M}_{\odot} $.
This dissipative process causes 
that fluctuations on scales below the Silk mass are completely washed out
at the time of recombination and no structure can form on these scales. 
This has consequences on large scale structure power spectrum, 
where small scales have very little power.

\begin{figure}[h]
  \begin{center}
    \leavevmode
    \epsfxsize = 10cm
    \epsffile{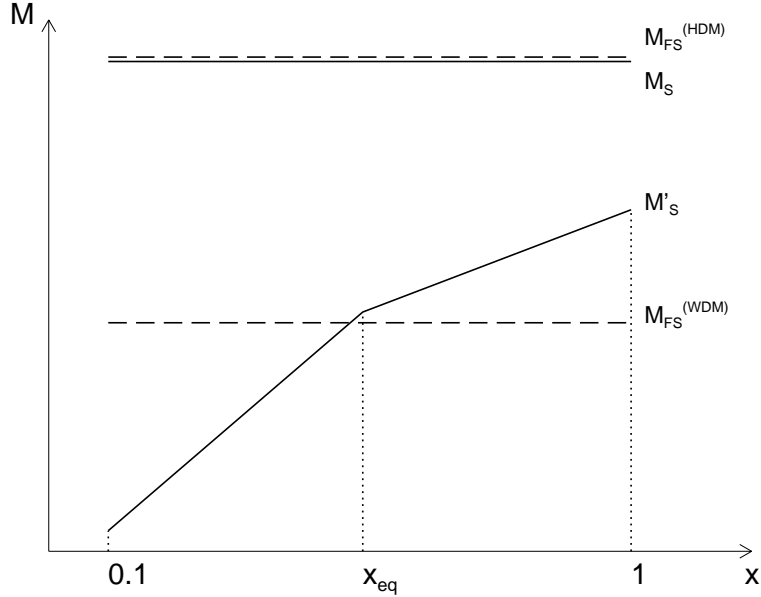}
  \end{center}
\caption{\small The trend of the mirror Silk mass ($ M'_{\rm S} $) over a 
cosmologically interesting range of $ x $, which contains $ x_{eq} $ (we 
considered $ \Omega_m h^2 \simeq 0.15 $, so 
$ x_{\rm eq} \simeq 0.3 $). 
The axis are both logarithmic. 
We show for comparison also the values of the ordinary Silk mass 
($ M_{\rm S} $) and of the free streaming mass ($ M_{\rm FS} $) for 
typical HDM and WDM candidates.}
\label{mirsilksca1}
\end{figure}

In the mirror sector too, obviously, the photon diffusion from the overdense 
to underdense regions induces a dragging of charged particles and washes 
out the perturbations at scales smaller than the mirror Silk scale 
\be
\lambda'_{\rm S} \simeq 3 f(x)(\beta \, \Omega_b h^2)^{-3/4} \;{\rm Mpc} \;, 
\label{mirsilklamb1}
\ee
where $f(x)=x^{5/4}$ for $x > x_{\rm eq}$ and 
$f(x) = (x/x_{\rm eq})^{3/2} x_{\rm eq}^{5/4}$ for $x < x_{\rm eq}$. 
Thus, the density perturbation scales running the linear growth after the 
matter-radiation equality epoch are limited by the length $ \lambda'_{\rm S} $. 
The smallest perturbations that survive the Silk damping will have the mass 
\be \label{ms_m}
M'_{\rm S} \sim [f(x) / 2]^3 (\beta \, \Omega_b h^2)^{-5/4} 10^{12}~ M_\odot \;,
\ee
which should be less than $ 2 \times 10^{12} ~ M_\odot $ in view of the BBN 
bound $ x < 0.64 $.  
Interestingly, for $ x \sim x_{\rm eq} $ we obtain, for the current estimate of 
$ \Omega_m h^2 $ and if all the dark matter is made of mirror baryons, 
$ M'_S \sim 10^{10} ~M_\odot $, a typical galaxy mass.

At this point it is very interesting a comparison between different damping 
scales, collisional (ordinary and mirror baryons) and collisionless 
(non-baryonic dark matter). 
We know that for hot dark matter (as a neutrino with mass $ \sim $10 eV) 
$ M_{\rm FS}^\nu \sim 10^{15} ~ M_\odot $, while for a typical warm dark 
matter candidate with mass $ \sim $1 keV, 
$ M_{\rm FS}^{WDM} \sim 10^{9} - 10^{10} ~ M_\odot $. 
From eq.~(\ref{ms_m}) it is evident that the dissipative scale for mirror 
Silk damping is analogous to that for WDM free streaming. 
Consequently, the cutoff effects on the corresponding large scale structure 
power spectra are similar, though with important differences due to the 
presence of oscillatory features, which makes 
them distinguishable one from the other (for a detailed presentation of these 
power spectra see the Paper 2 \cite{paper2}). 
In figure \ref{mirsilksca1} we show this comparison together with the trend of 
the mirror Silk mass over a cosmologically interesting range of $ x $.


\def \mirror_bar_struc{Scenarios}
\section{\mirror_bar_struc}
\label{mirror_bar_struc}

\begin{figure}[h]
  \begin{center}
    \leavevmode
    \epsfxsize = 10cm
    \epsffile{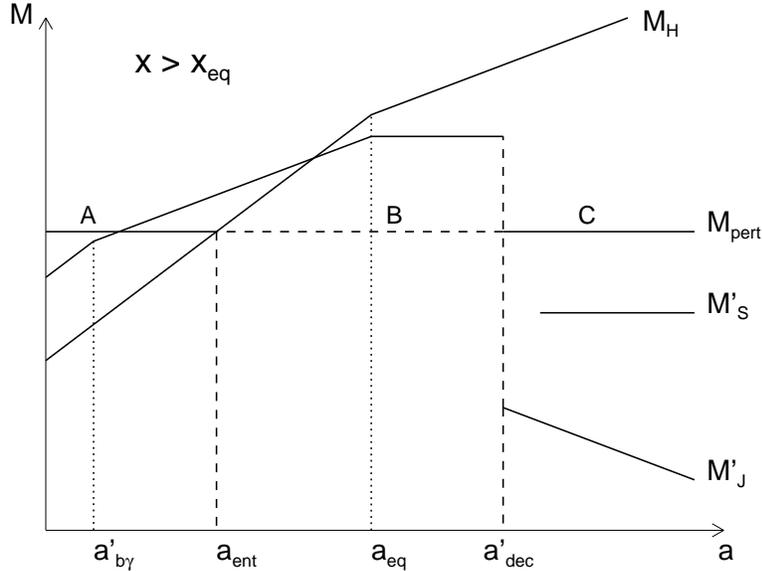}
  \end{center}
\caption{\small Typical evolution of a perturbed scale $ M_{\rm pert} $ 
in adiabatic mirror baryonic dark matter scenario with $ x > x_{\rm eq} $. 
The figure shows the Jeans mass $ M_{\rm J}' $, the Silk mass 
$ M_{\rm S}' $ and the Hubble mass $ M_{\rm H} $. 
The time of horizon crossing of the perturbation is indicated by 
$ a_{\rm ent} $. 
Are also indicated the three evolutionary stages: during stage A 
($ a < a_{\rm ent} < a_{\rm eq} $) the mode grows as $ \delta_b \propto a^2 $; 
throughout stage B ($ a_{\rm ent} < a < a_{\rm dec}' $) the perturbation 
oscillates; finally, in stage C ($ a > a_{\rm dec}' $) the mode becomes 
unstable again and grows as $ \delta_b \propto a $. 
Fluctuations with size smaller than $ M_{\rm S}' $ are wiped out 
by photon diffusion.}
\label{mirmasssca1}
\end{figure}

\begin{figure}[h]
  \begin{center}
    \leavevmode
    \epsfxsize = 10cm
    \epsffile{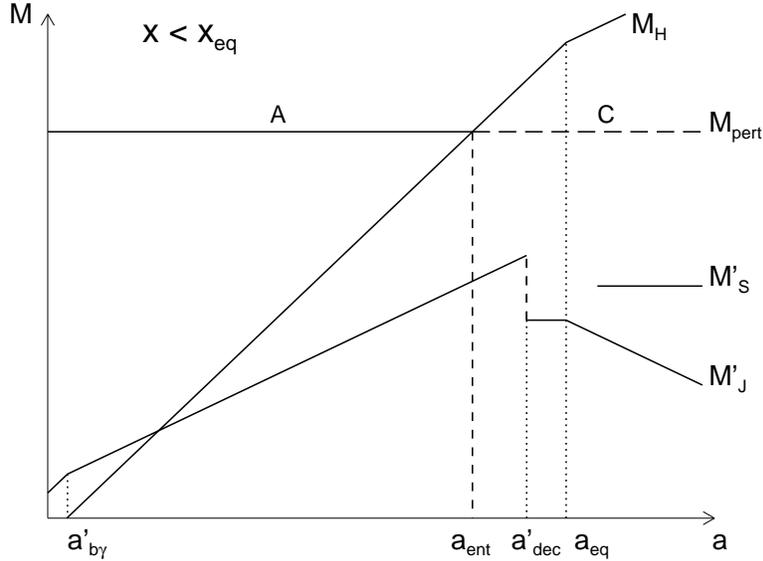}
  \end{center}
\caption{\small Typical evolution of a perturbed scale $ M_{\rm pert} $ 
in adiabatic mirror baryonic dark matter scenario with $ x < x_{\rm eq} $. 
The value of $ M_{\rm pert} $ is the same as in figure \ref{mirmasssca1}. 
The time of horizon crossing of the perturbation is indicated by 
$ a_{\rm ent} $. 
The figure shows the Jeans mass $ M_{\rm J}' $, the Silk mass 
$ M_{\rm S}' $ and the Hubble mass $ M_{\rm H} $. 
Unlike the case $ x > x_{\rm eq} $ (shown in the previous figure), now there 
are only the two evolutionary stages A ($ a < a_{\rm ent} $) and C 
($ a > a_{\rm ent} $). 
Fluctuations with size smaller than $ M_{\rm S}' $ are wiped out by photon 
diffusion, but in this case the Silk mass is near to the maximum 
Jeans mass.}
\label{mirmasssca2}
\end{figure}

After the description of the fundamental scales for structure formation, let us 
now collect all the informations and discuss the mirror scenarios. 
They are essentially two, according to the value of $ x $, which can be 
higher or lower than $ x_{\rm eq} $, and are shown respectively in figures 
\ref {mirmasssca1} and \ref {mirmasssca2}, which will be our references 
during the present section.

Typically, adiabatic perturbations for mirror baryons with sizes larger than 
the maximum value of the Jeans mass, which is $ M_{\rm J}'(a_{\rm eq}) $ 
for $ x > x_{\rm eq} $ and $ M_{\rm J}'(a_{\rm dec}') $ for $ x < x_{\rm eq} $, 
experience uninterrupted growth. 
In particular, they grow as $ \delta_b \propto a^2 $ before matter-radiation 
equality and as $ \delta_b \propto a $ after equality. 
Fluctuations on scales in the mass interval 
$ M_{\rm S}' < M < M_{\rm J,max} $ grow as $ \delta_b \propto a^2 $ 
while they are still outside the Hubble radius. 
After entering the horizon and until recombination these modes oscillate like 
acoustic waves. 
The amplitude of the oscillation is constant before equilibrium but 
decreases as $ a^{-1/4} $ between equipartition and recombination. 
After decoupling the modes become unstable again and grow as 
$ \delta_b \propto a $. 
Finally all perturbations on scales smaller than the value of the Silk mass 
are dissipated by photon diffusion. 

Given this general behaviour, the schematic evolution of an adiabatic mode 
with a reference mass scale $ M_{\rm pert} $, with 
$ M_{\rm S}' < M_{\rm pert} < M_{\rm J}'(a_{\rm eq}) $, is depicted in figure 
\ref{mirmasssca1} for $ x > x_{\rm eq} $.
We distinguish between three evolutionary stages, called A, B and C, 
depending on the size of the perturbation and 
on the cosmological parameters $ \Omega_b h^2 $, $ x $ and $ \beta $, 
which determine the behaviour of the mass scales, and in particular the key 
moments (time of horizon crossing and decoupling) and the dissipative Silk 
scale. 
During stage A, i.e. before the horizon crossing 
($ a < a_{\rm ent} < a_{\rm eq} $), the mode grows as 
$ \delta_b \propto a^2 $; throughout stage B 
($ a_{\rm ent} < a < a_{\rm dec}' $) the perturbation enters the horizon, 
baryons and photons feel each other, and it oscillates; finally, in stage C 
($ a > a_{\rm dec}' $), the photons and baryons decouple, and the mode 
becomes unstable again growing as $ \delta_b \propto a $. 
We remark that fluctuations with sizes greater than $M_{\rm J}'(a_{\rm eq})$ 
grow uninterruptedly (because after horizon crossing the photon pressure 
cannot balance the gravity), changing the trend from $ a^2 $ before MRE to 
$ a $ after it, while those with sizes smaller than $ M_{\rm S}' $ are 
completely washed out by photon diffusion.

After decoupling, all surviving perturbations (those with 
$ M_{\rm pert} > M'_{\rm S} $) grow steadily until their amplitude becomes 
so large that the linear theory breaks down and one needs to employ a 
different type of analysis. 

If we look, instead, at the schematic evolution of an adiabatic mode with the 
same reference mass scale $ M_{\rm pert} $ but for $ x < x_{\rm eq} $, as 
reported in figure \ref{mirmasssca2}, we immediately notice the lower 
values of the maximum Jeans mass and the Silk mass, which are similar. 
Therefore, for the plotted perturbative scale there are now only the two 
stages A and C. 
In general, depending on its size, the perturbation mass can be higher or 
lower than the Silk mass (and approximatively also than the maximum 
Jeans mass), so modes with $ M_{\rm pert} > M'_{\rm S} $ grow 
continuously before and after their horizon entry, while modes with 
$ M_{\rm pert} < M'_{\rm S} $ are completely washed out.

We find that $ M'_{\rm J} $ becomes smaller than the Hubble horizon mass 
$ M_{\rm H} $ starting from a redshift 
\be 
z_{\rm c} = 3750 \: x^{-4} \: (\Omega_m h^2) \;, 
\ee
which is about $ z_{\rm eq} $ for $ x=0.64 $, but it sharply increases for 
smaller values of $ x $, as shown in figure \ref{figzz}. 
We can recognize this behaviour also watching at the intersections of the 
lines for $ M'_{\rm J} $ and $ M_{\rm H} $ in figures \ref{mirmasssca1} and 
\ref{mirmasssca2}. 
Thus, density perturbation scales which enter horizon at 
$ z \sim z_{\rm eq} $ have masses larger than $ M'_J $ and thus undergo 
uninterrupted linear growth immediately after $ t_{\rm eq} $. 
Smaller scales for which $ M'_{\rm J} > M_{\rm H} $ would instead first 
oscillate. 
Therefore, the large scale structure formation is not delayed even if the 
mirror decoupling did not occur yet, i.e. even if $ x > x_{\rm eq} $. 

When compared with non baryonic dark matter scenarios, 
the main feature of the mirror baryonic scenario is that the M 
baryon density fluctuations should undergo the strong collisional damping 
around the time of M recombination, which washes out the perturbations 
at scales smaller than the mirror Silk scale. 
It follows that density perturbation scales which undergo the linear growth 
after the MRE epoch are limited by the length $ \lambda'_{\rm S}$. 
This could help in avoiding the excess of small scales (of few Mpc) in the 
CDM power spectrum without tilting the spectral index.
To some extent, the cutoff effect is analogous to the free streaming damping 
in the case of warm dark matter (WDM), but there are important differences. 
The point is that, alike usual baryons, the MBDM shows acoustic 
oscillations with an impact on the large scale structure (LSS) power 
spectrum \cite{paolo,paper2}. 
In particular, it is tempting to imagine that the M baryon oscillation effects 
are related to the anomalous features observed in LSS power spectra.  

In addition, the MBDM oscillations transmitted via gravity to the ordinary 
baryons, could cause observable anomalies in the CMB angular power 
spectrum for $ l $'s larger than 200. 
This effect can be observed only if the M baryon Jeans scale 
$ \lambda'_{\rm J} $ is larger than the Silk scale of ordinary baryons, 
which sets a principal cutoff for CMB oscillations around $ l \sim 1200 $. 
This would require enough large values 
of $ x $, near the upper bound fixed by the BBN constraints, 
and, together with the possible effects on the large scale power spectrum, it 
can provide a direct test for the MBDM (verifiable by the higher sensitivity 
of next CMB and LSS observations).  
For a complete discussion of the CMB and LSS power spectra for a Mirror 
Universe see the Paper 2 \cite{paper2}.

{\em Clearly, for small $ x $ the M matter recombines before the MRE 
moment, and thus it behavesas the CDM as far as the large scale 
structure is concerned.} 
However, there still can be crucial differences at smaller scales which 
already went non-linear, like galaxies. 
In our scenario, dark matter in galaxies and clusters can contain both CDM 
and MBDM components, or can be even constituted entirely by the mirror 
baryons. 

One can question whether the MBDM distribution in halos can be different 
from that of the CDM. 
Simulations show that the CDM forms triaxial halos with a density profile too 
clumped toward the center, and overproduces the small substructures 
within the halo.
Since MBDM constitutes a kind of collisional dark matter, it may potentially 
avoid these problems, at least the one related with the excess of small 
substructures. 

It's also worth noting that, throughout the above discussion, we have 
assumed that the matter density of the Universe is close to unity. 
If, instead, the matter density is small and a vacuum density contribution is 
present, there is an additional complication due to the fact that the Universe 
may become curvature dominated starting from some redshift 
$ z_{\rm curv} $. 
Given the current estimate $ \Omega_\Lambda \simeq 0.7 $, 
this transition has yet occurred and the growth of perturbations has stopped 
around $ z_{\rm curv} $, when the expansion became too rapid for it.


\def \mirevolpert{Evolution of perturbations}
\section{\mirevolpert}
\label{mirevolpert}

As a result of the studies done in previous sections, in this section we 
finally consider the temporal evolution of perturbations, as function of the 
scale factor $ a $. All the plots are the results of numerical computations 
obtained using a Fortran code originally written for the ordinary 
Universe and modified to account for the mirror sector.

We used the synchronous gauge and the evolutionary equations presented 
in ref.~\cite{mabert}. 
The difference in the use of other gauges is limited to the gauge-dependent 
behaviour of the density fluctuations on scales larger than the horizon. 
The fluctuations can appear as growing modes in one coordinate system 
and as constant mode in another, that is exactly what occurs in the 
synchronous and the conformal Newtonian gauges.

In the figures we plot the evolution of the components of a Mirror Universe, 
namely the cold dark matter\footnote{
As non baryonic dark matter we consider only the cold dark matter, which is 
at present the standard choice in cosmology.
}, the ordinary baryons and photons, and the mirror baryons and photons, 
changing some parameter to evaluate their influence. 

\begin{figure}[p]
\begin{center}
\leavevmode
{\hbox 
{\epsfxsize = 12.5cm \epsfysize = 9.5cm 
    \epsffile{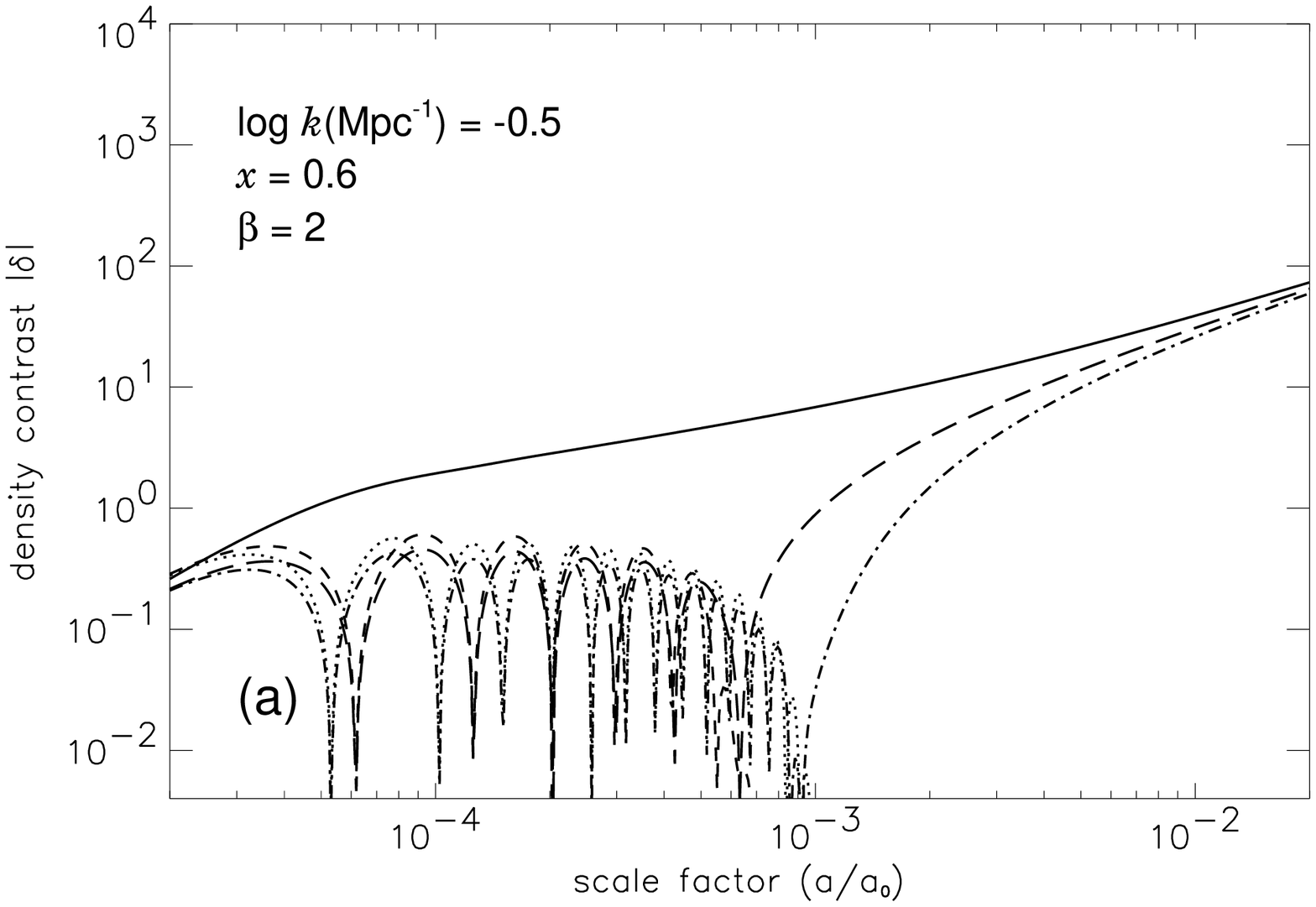} }
{\epsfxsize = 12.5cm \epsfysize = 9.5cm 
    \epsffile{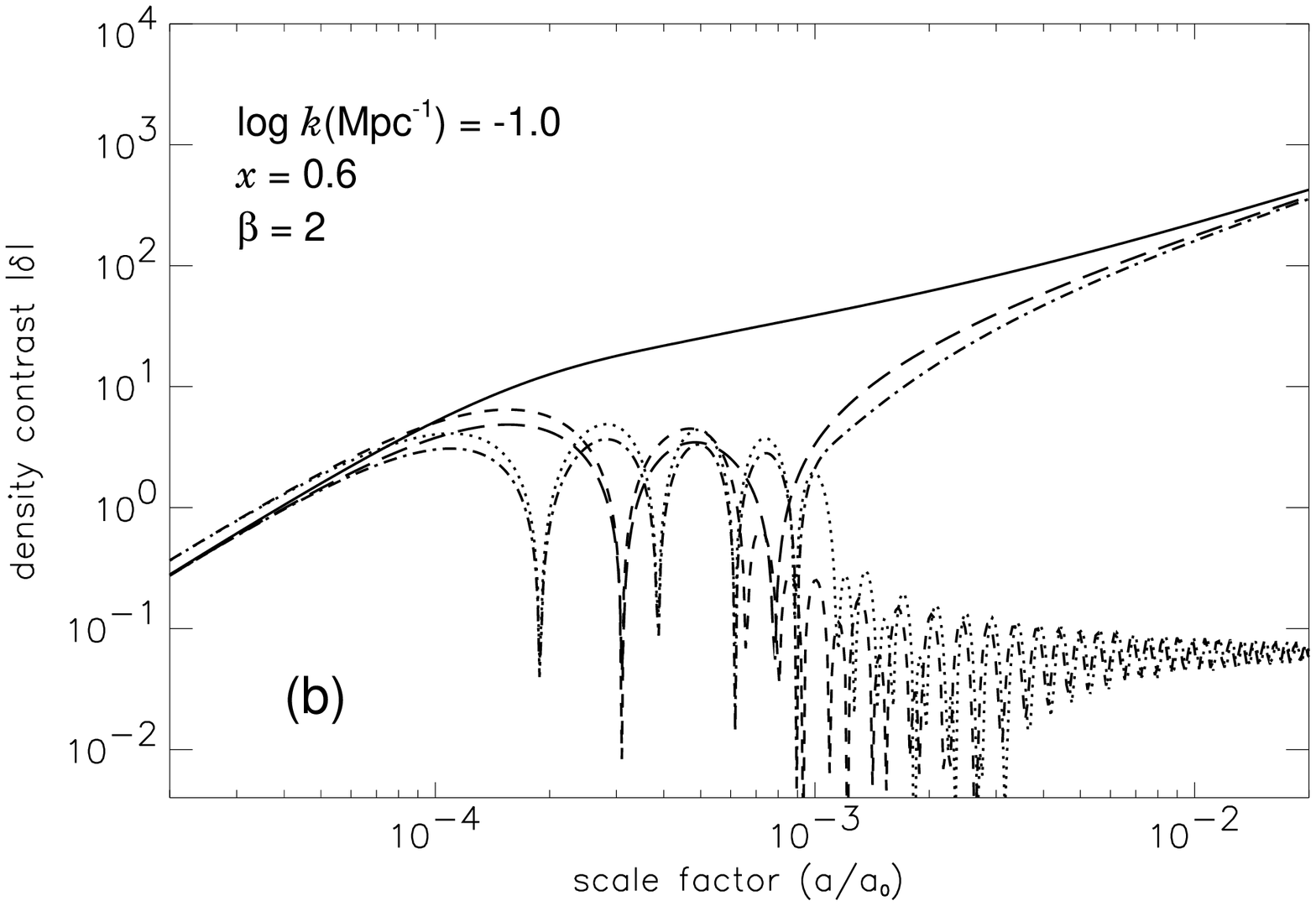} } }
\end{center}
\vspace{-.5cm}
\caption{\small Evolution of perturbations for the components of a Mirror 
Universe: cold dark matter (solid line), ordinary baryons and photons 
(dot-dashed and dotted) and mirror baryons and photons 
(long dashed and dashed). 
The model is a flat Universe with $ \Omega_m = 0.3 $, 
$ \Omega_b h^2 = 0.02 $, $ \Omega'_b h^2 = 0.04 $ ($ \beta = 2 $), 
$ h = 0.7 $, $ x = 0.6 $, and plotted scales are 
$ \log k ({\rm Mpc}^{-1}) = -0.5 $  ($ a $) and $ -1.0 $ ($ b $).}
\label{evol-x06-b2-k0510}
\end{figure}

First of all, we comment figure \ref{evol-x06-b2-k0510}b, which is the most 
useful to recognize the general features of the evolution of perturbations. 
Starting from the smallest scale factor, we see that all three matter 
components and the two radiative components grow with the same trend 
(as $ a^2 $), but the radiative ones have a slightly higher density contrast 
(with a constant rate until they are tightly coupled); this is simply the 
consequence of considering adiabatic perturbations, which are linked in their 
matter and radiation components by the adiabatic condition (\ref{adcond}). 
This is the situation when the perturbation is out of horizon, but, when it 
crosses the horizon, around $ a \sim 10^{-4} $, things drastically change. 
Baryons and photons, in each sector separately, become causally 
connected, feel each other, and begin to oscillate for the competitive effects 
of gravity and pressure. 
Meanwhile, the CDM density perturbation continues to grow uninterruptedly, 
at first reducing his rate from $ a^2 $ to $ \ln a $ (due to the rapid expansion 
during the radiation era), and later, as soon as MRE occurs (at 
$ a \sim 3 \times 10^{-3} $ for the considered model), increasing 
proportionally to $ a $. 
The oscillations of baryons and photons continue until their decoupling, 
which in the mirror sector occurs before than in the ordinary one (scaled by 
the factor $ x $). 
This moment is marked in the plot as the point where the lines for the two 
components move away one from the other. 
From this point, the photons in both sectors continue the oscillations until 
they are completely damped, while the M and O baryons rapidly fall into 
the potential wells created by the cold dark matter and start growing as $ a $. 
We remark that it's important the way in which the oscillation reaches the 
decoupling; if it is compressing, first it slows down (as if it continues to 
oscillate, but disconnected from the photons), and then it expands driven 
by the gravity; if, otherwise, it is expanding, it directly continues the 
expansion and we see in the plot that it immediately stops to oscillate. 
In this figure we have the first behaviour in the mirror sector, the second one 
in the ordinary sector.

\begin{figure}[p]
\begin{center}
\leavevmode
{\hbox 
{\epsfxsize = 12.5cm \epsfysize = 9.5cm 
    \epsffile{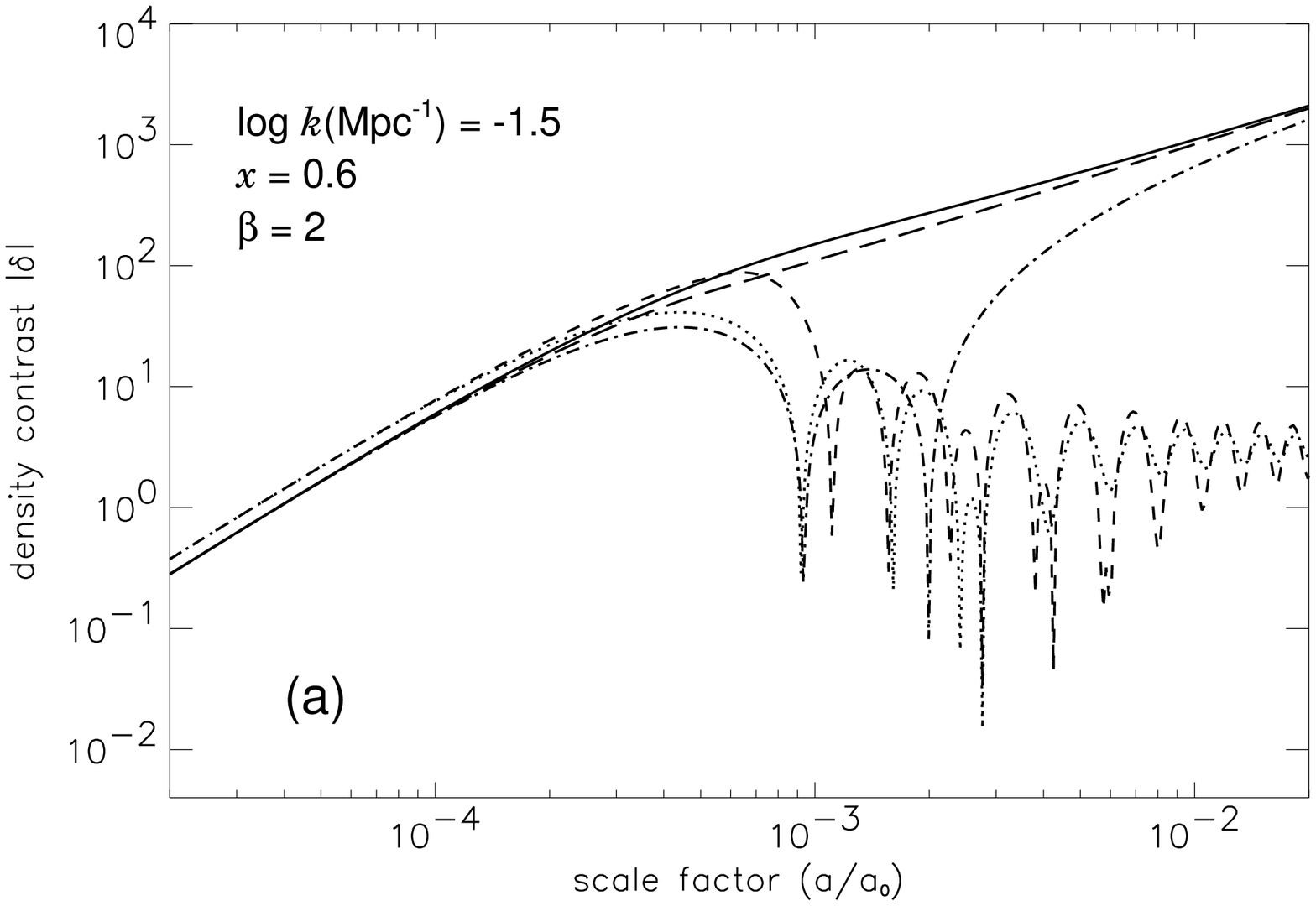} }
{\epsfxsize = 12.5cm \epsfysize = 9.5cm 
    \epsffile{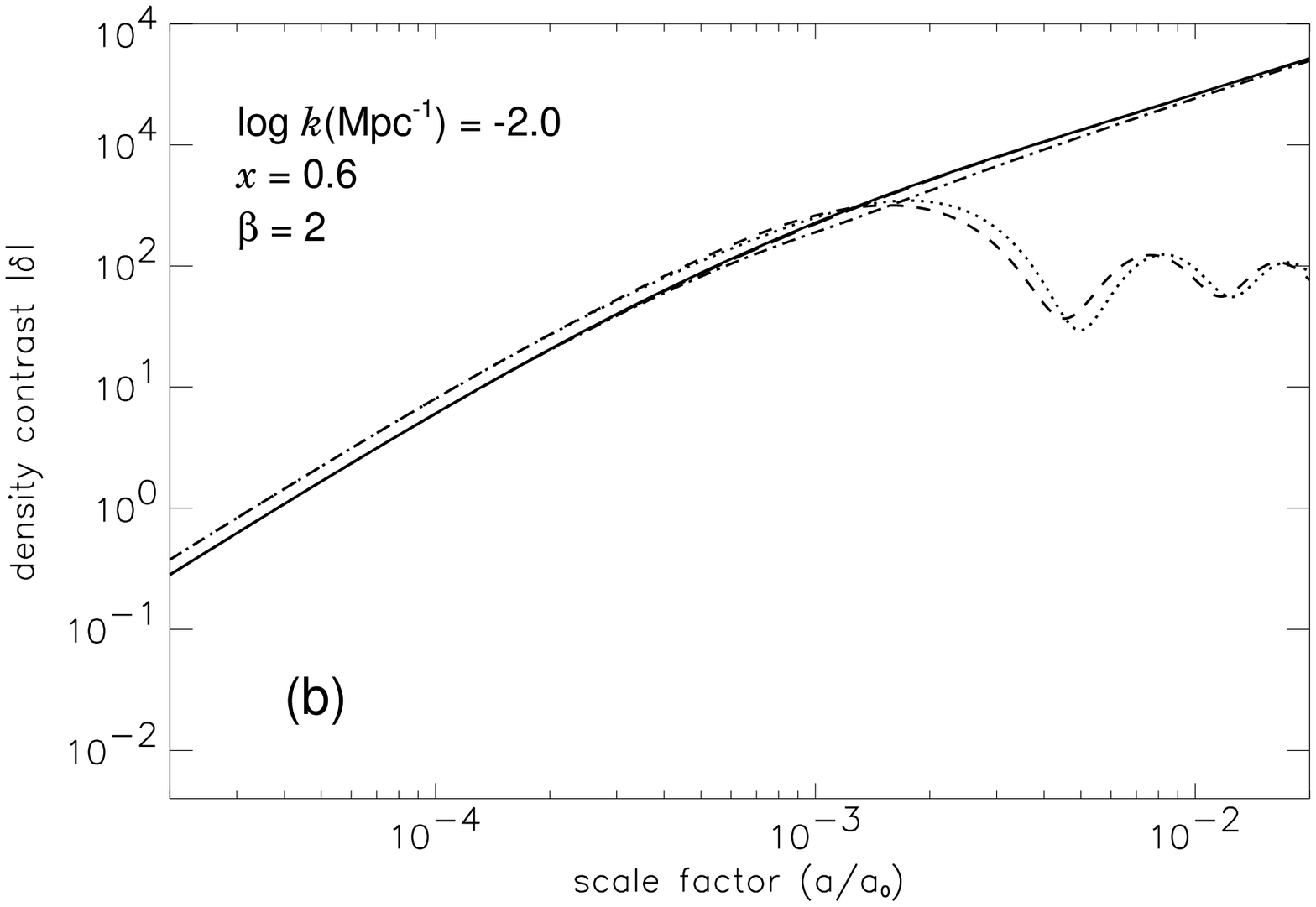} } }
\end{center}
\caption{\small The same as in figure \ref{evol-x06-b2-k0510}, but for scales 
$ \log k ({\rm Mpc}^{-1}) = -1.5 $  ($ a $) and $ -2.0 $ ($ b $).}
\label{evol-x06-b2-k1520}
\end{figure}

\begin{figure}[p]
\begin{center}
\leavevmode
{\hbox 
{\epsfxsize = 12.5cm \epsfysize = 9.5cm 
    \epsffile{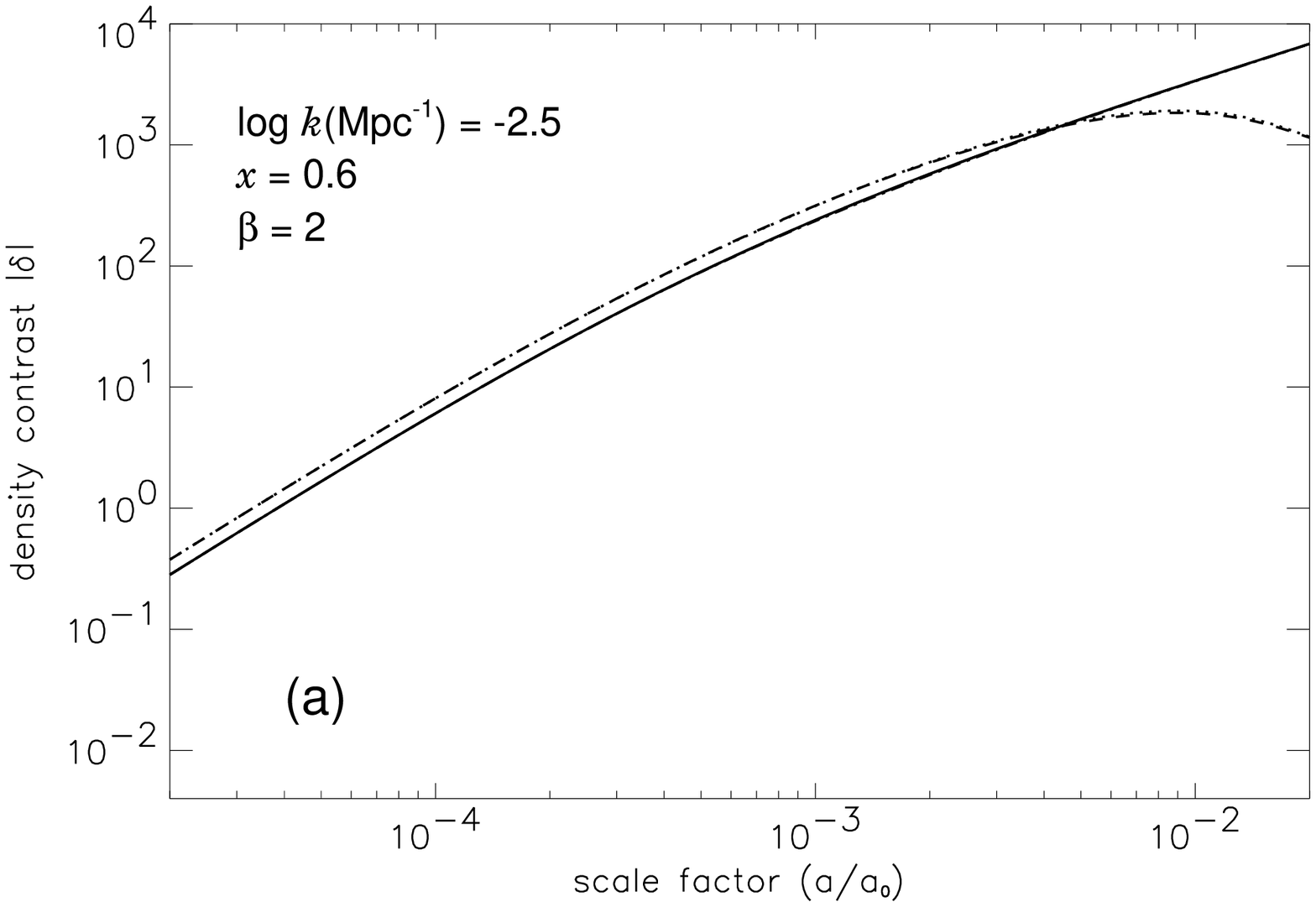} }
{\epsfxsize = 12.5cm \epsfysize = 9.5cm 
    \epsffile{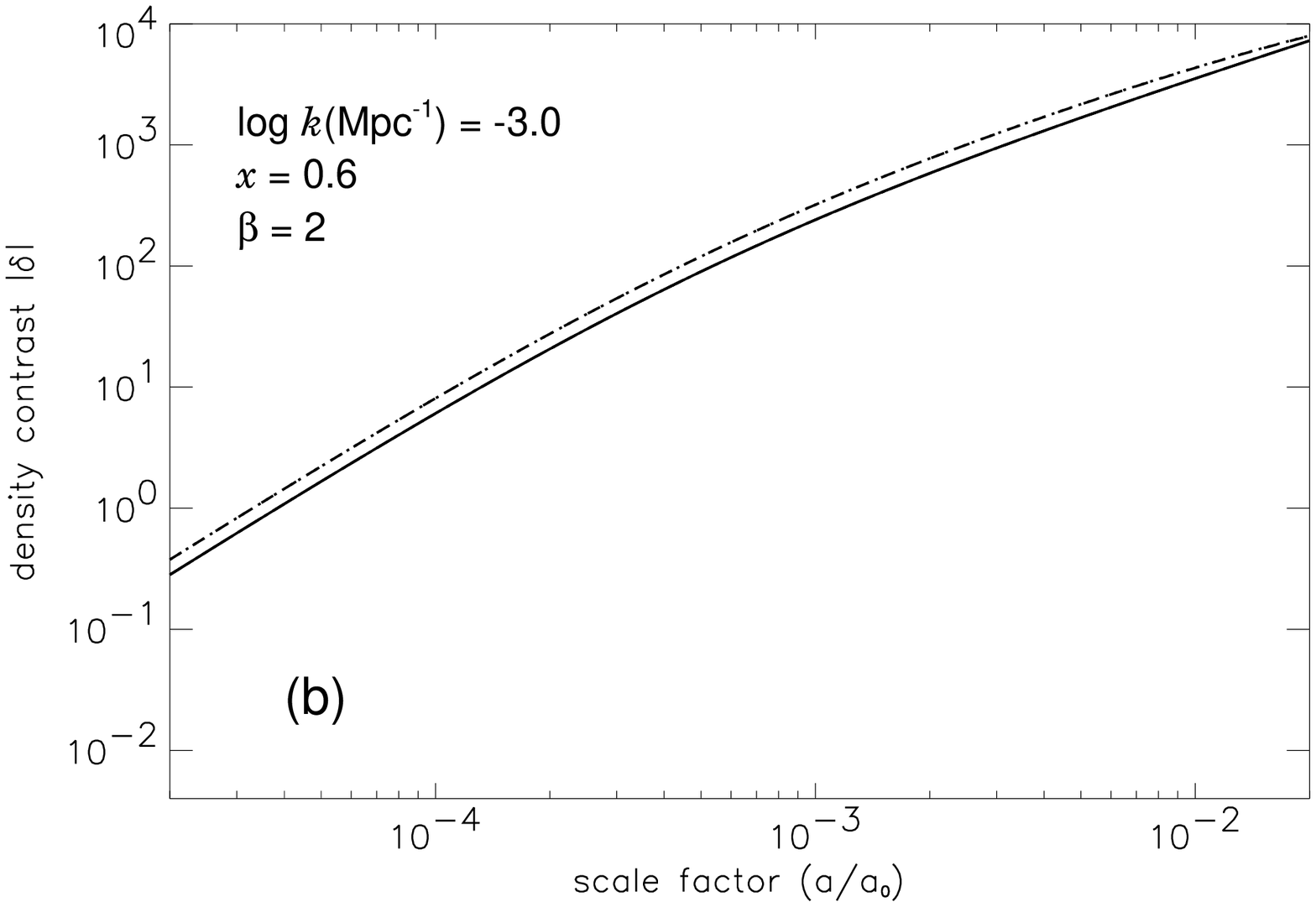} } }
\end{center}
\caption{\small The same as in figure \ref{evol-x06-b2-k0510}, but for scales 
$ \log k ({\rm Mpc}^{-1}) = -2.5 $  ($ a $) and $ -3.0 $ ($ b $).}
\label{evol-x06-b2-k2530}
\end{figure}

In figures \ref{evol-x06-b2-k0510}, \ref{evol-x06-b2-k1520} and 
\ref{evol-x06-b2-k2530} we compare the behaviours of different scales for 
the same model. 
The scales are given by 
$ \log k (\rm Mpc^{-1}) = -0.5, -1.0, -1.5, -2.0, -2.5, -3.0 $, where 
$ k = 2 \pi / \lambda $ is the wave number. 
The effect of early entering the horizon of small scales (those with higher 
$ k $) is evident. 
Going toward bigger scales the superhorizon growth continue for a longer 
time, delaying more and more the beginning of acoustic oscillations, until it 
occurs out of the coordinate box for the bigger plotted scale ($ \log k = -3.0 $).
Starting from the scale $ \log k = -1.5 $, the mirror decoupling occurs before 
the horizon entry of the perturbations, and the evolution of the mirror 
baryons density is similar to that of the CDM. 
The same happens to the ordinary baryons too, but for $ \log k \lsim -2.0 $ 
(since they decouple later), while the evolution of mirror baryons is yet 
indistinguishable from that of the CDM. 
For the bigger scales ($ \log k \lsim -2.5 $) the evolution of all three matter 
components is identical.

\begin{figure}[p]
\begin{center}
\leavevmode
{\hbox 
{\epsfxsize = 12.5cm \epsfysize = 9.5cm 
    \epsffile{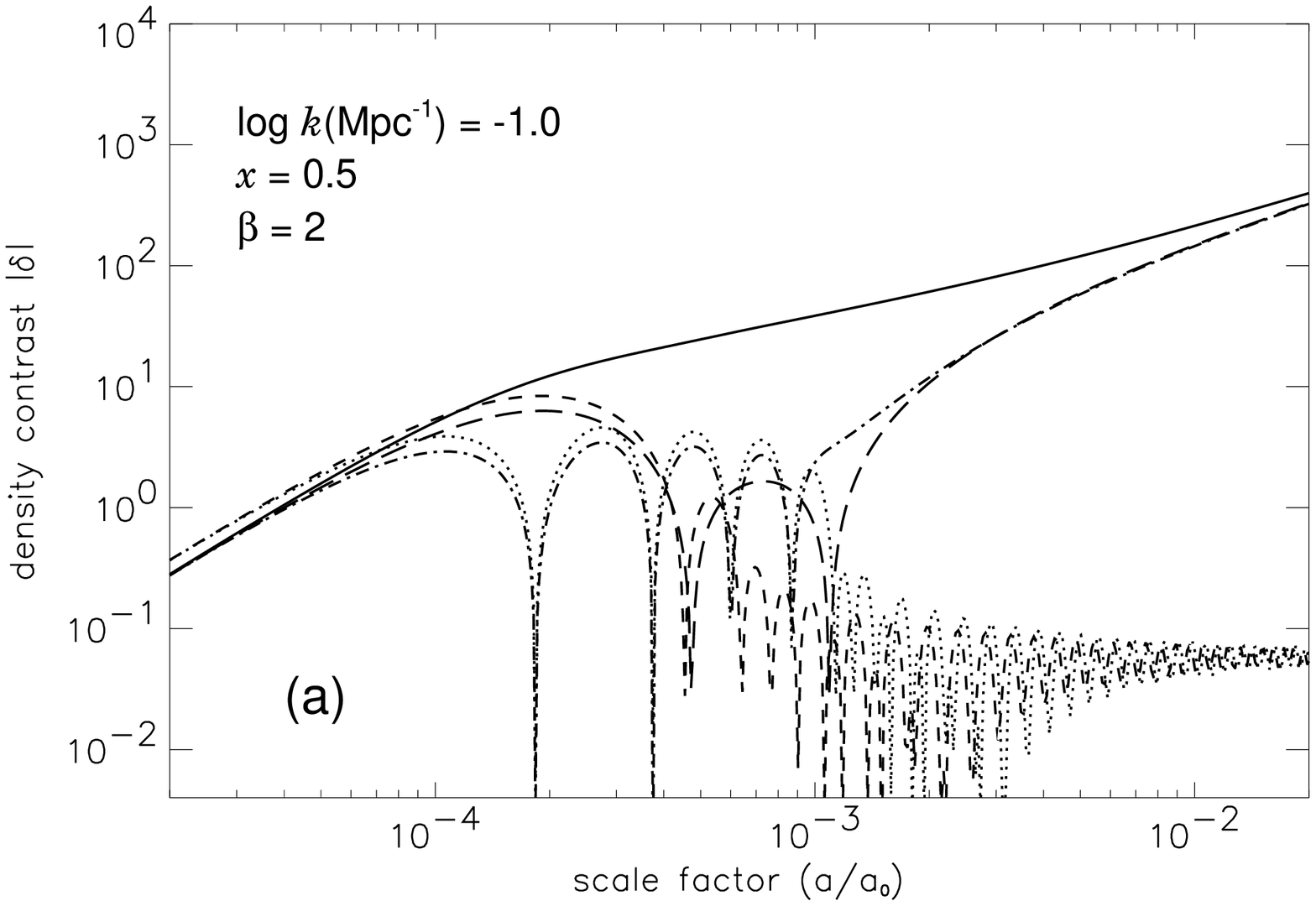} }
{\epsfxsize = 12.5cm \epsfysize = 9.5cm 
    \epsffile{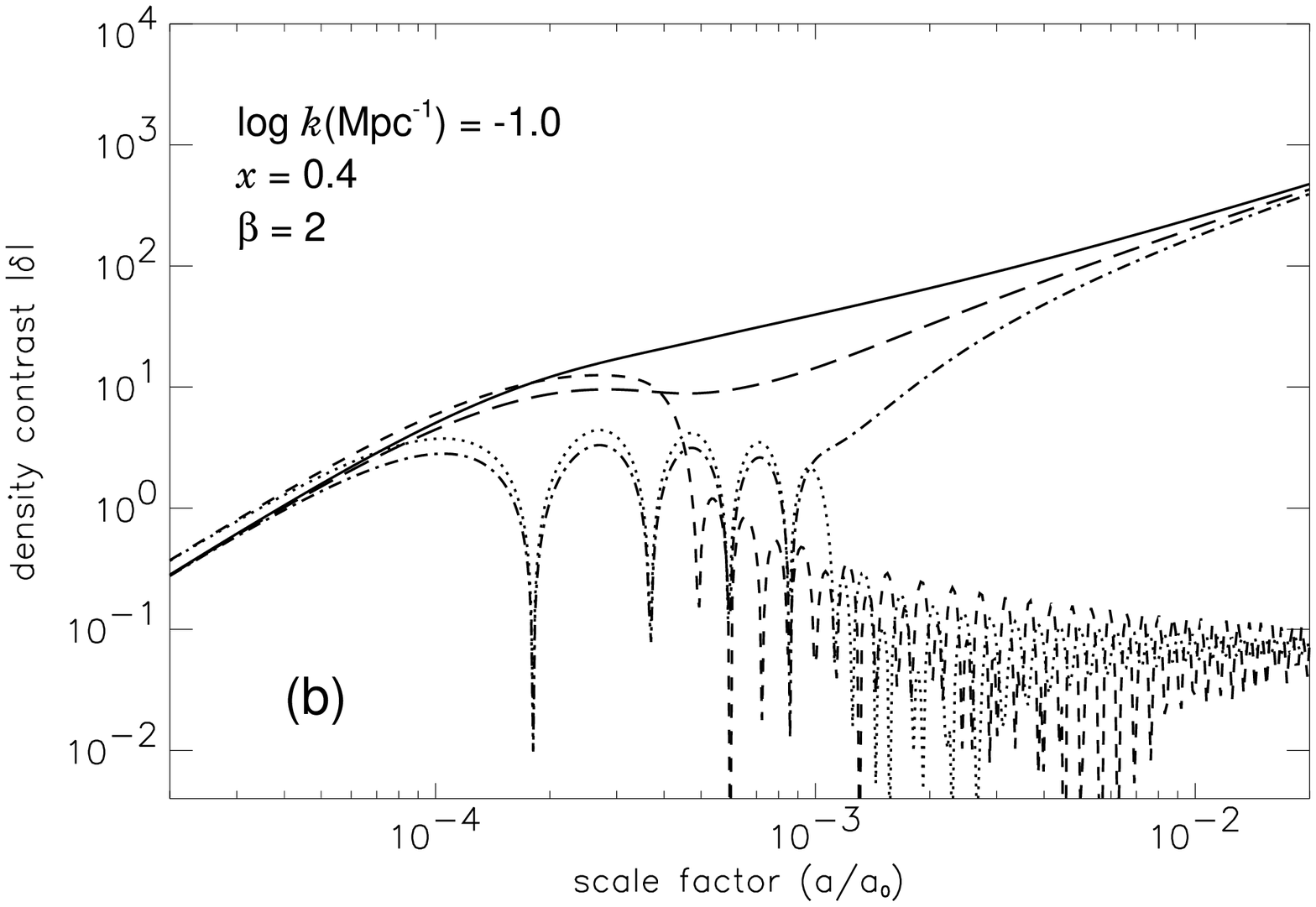} } }
\end{center}
\caption{\small Evolution of perturbations for the components of a Mirror 
Universe: cold dark matter (solid line), ordinary baryons and photons 
(dot-dashed and dotted) and mirror baryons and photons 
(long dashed and dashed). 
The model is a flat Universe with $ \Omega_m = 0.3 $, 
$ \Omega_b h^2 = 0.02 $, $ \Omega'_b h^2 = 0.04 $ ($ \beta = 2 $), 
$ h = 0.7 $, $ x = 0.5 $ ($ a $) or $ 0.4 $ ($ b $), and plotted scale is 
$ \log k ({\rm Mpc}^{-1}) = -1.0 $.}
\label{evol-x0504-b2-k10}
\end{figure}

\begin{figure}[p]
\begin{center}
\leavevmode
{\hbox 
{\epsfxsize = 12.5cm \epsfysize = 9.5cm 
    \epsffile{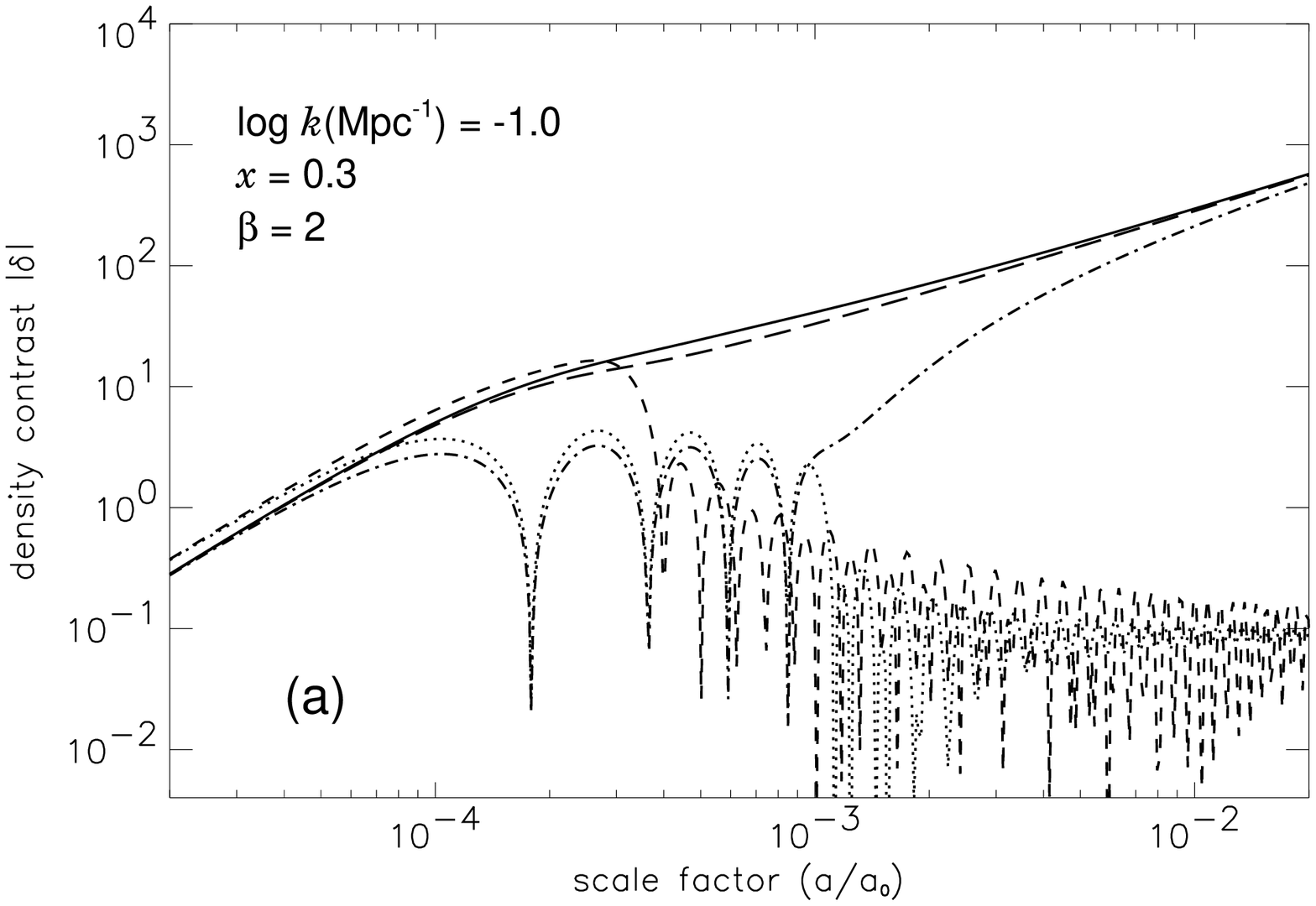} }
{\epsfxsize = 12.5cm \epsfysize = 9.5cm 
    \epsffile{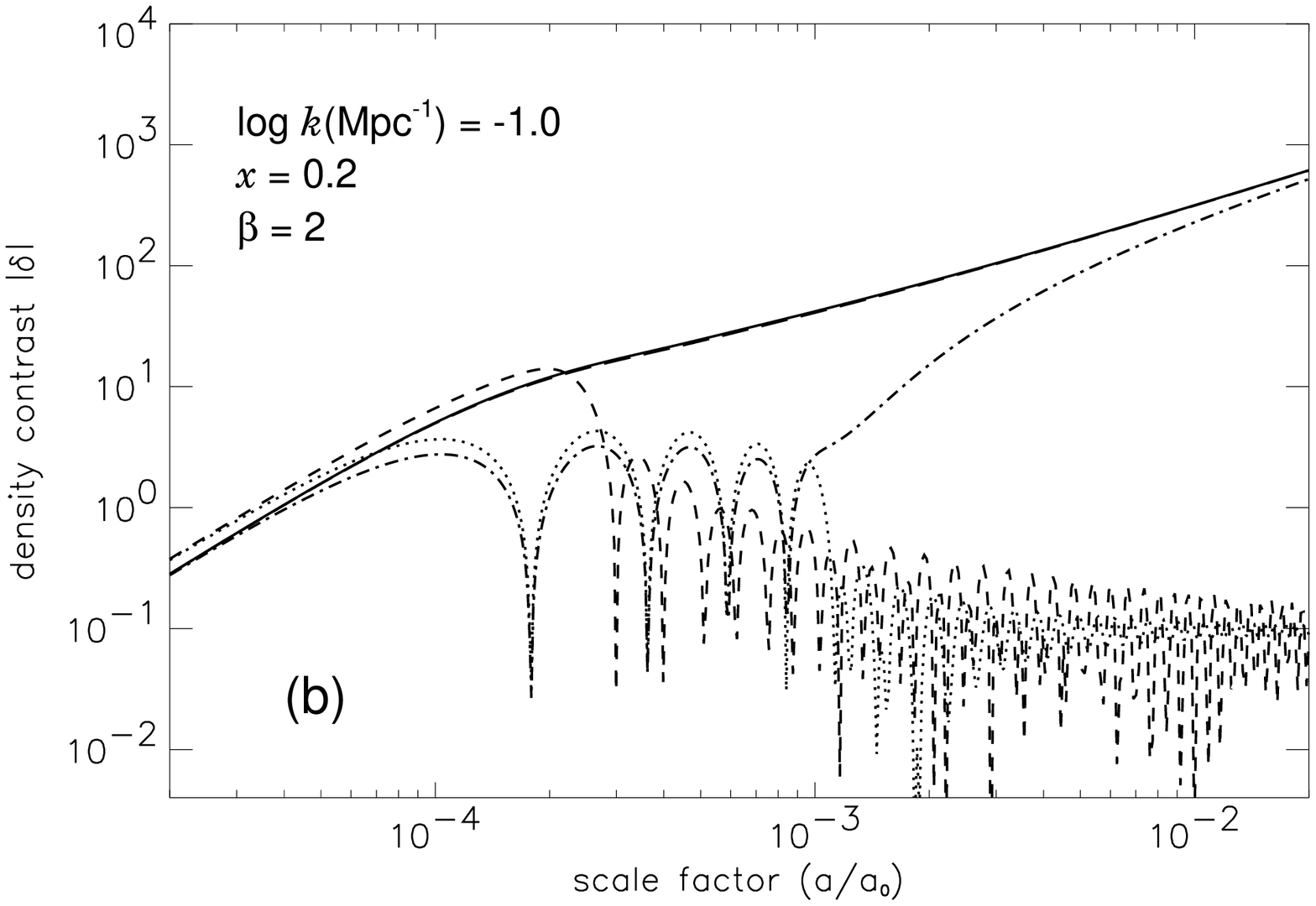} } }
\end{center}
\caption{\small The same as in figure \ref{evol-x0504-b2-k10}, but for 
$ x = 0.3 $ ($ a $) and $ 0.2 $ ($ b $).}
\label{evol-x0302-b2-k10}
\end{figure}

\begin{figure}[p]
  \begin{center}
    \leavevmode
    \epsfxsize = 12.5cm
    \epsfysize = 9.5cm
    \epsffile{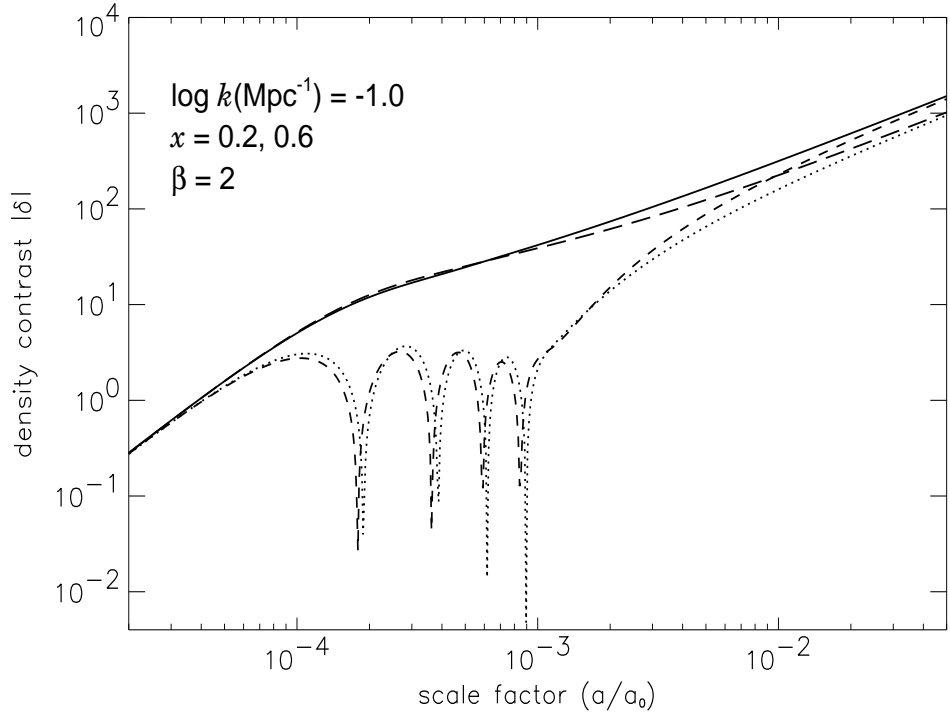}
  \end{center}
\caption{\small Evolution of perturbations in a Mirror Universe for 
cold dark matter (solid and long dashed lines) and ordinary baryons 
(dashed and dotted). 
The models are flat, with $ \Omega_m = 0.3 $, 
$ \Omega_b h^2 = 0.02 $, $ \Omega'_b h^2 = 0.04 $ ($ \beta = 2 $), 
$ h = 0.7 $, $ x = 0.2 $ (solid and dashed) and $ 0.6 $ (long dashed 
and dotted), and $ \log k ({\rm Mpc}^{-1}) = -1.0 $.
The highest scale factor plotted here is 0.05, while in all other 
figures it is 0.02.}
\label{evol-x0206-b2-k10}
\end{figure}

As previously seen, the decoupling is a crucial point for structure formation, 
and it assumes a fundamental role specially in the mirror sector, where it 
occurs before than in the ordinary one: mirror baryons can start before 
growing perturbations in their density distribution. 
For this reason it's important to analyze the effect of changing the mirror 
decoupling time, obtained changing the value of $ x $ and leaving 
unchanged all other parameters, as it is possible to do using figures 
\ref{evol-x06-b2-k0510}b, \ref{evol-x0504-b2-k10} and 
\ref{evol-x0302-b2-k10} for $ x = 0.6, 0.5, 0.4, 0.3, 0.2 $ and the same scale 
$ \log k = -1.0 $. 
It is evident the shift of the mirror decoupling toward lower values of $ a $ 
when reducing $ x $, according to the law (\ref{z'_dec}), which states a direct 
proportionality between the two. 
In particular, for $ x < x_{\rm eq} \approx 0.3 $ mirror decoupling occurs 
before the horizon crossing of the perturbation, and mirror baryons mimic 
more and more the CDM, so that for $ x \simeq 0.2 $ the perturbations in the 
two components are indistinguishable. 
For the ordinary sector apparently there are no changes, but at a more 
careful inspection we observe some difference due to the different amount of 
relativistic mirror species (proportional to $ x^4 $), which slightly shifts the 
matter-radiation equality. 
This effect is more clear in figure \ref{evol-x0206-b2-k10}, where we plot 
only the CDM and the ordinary baryons for the cases $ x = 0.2 $ and $ 0.6 $: 
for the lower value of $ x $ there are less mirror photons, the MRE occurs 
before and the perturbation in the collisionless component starts growing 
before proportionally to the scale factor; thus, when the baryons decouple, 
their perturbation rapidly grows to equalize that in the CDM, which 
meanwhile has raised more for the lower $ x $.

\begin{figure}[p]
  \begin{center}
    \leavevmode
{\hbox 
{\epsfxsize = 12.5cm \epsfysize = 9.5cm 
    \epsffile{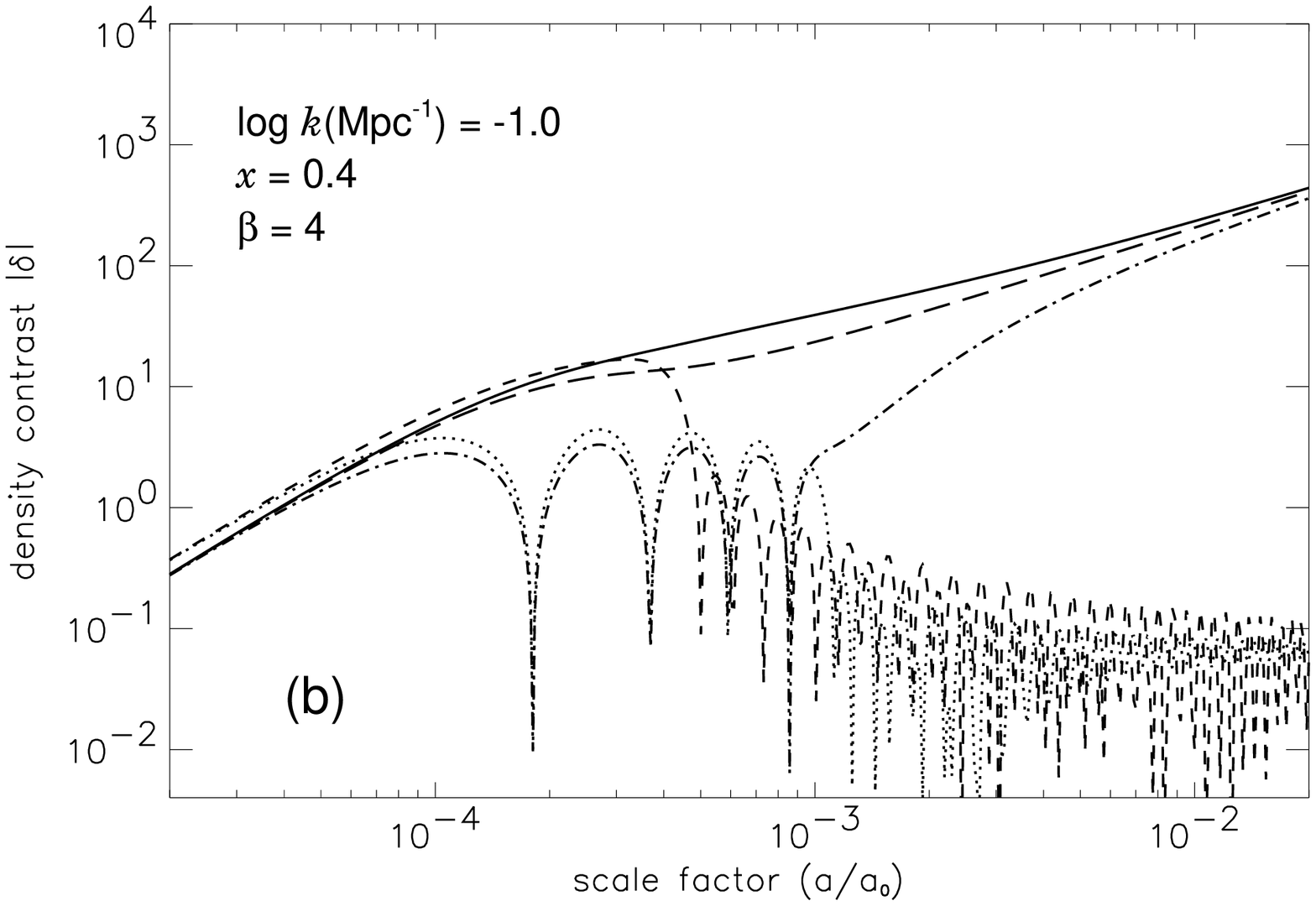} }
{\epsfxsize = 12.5cm \epsfysize = 9.5cm 
    \epsffile{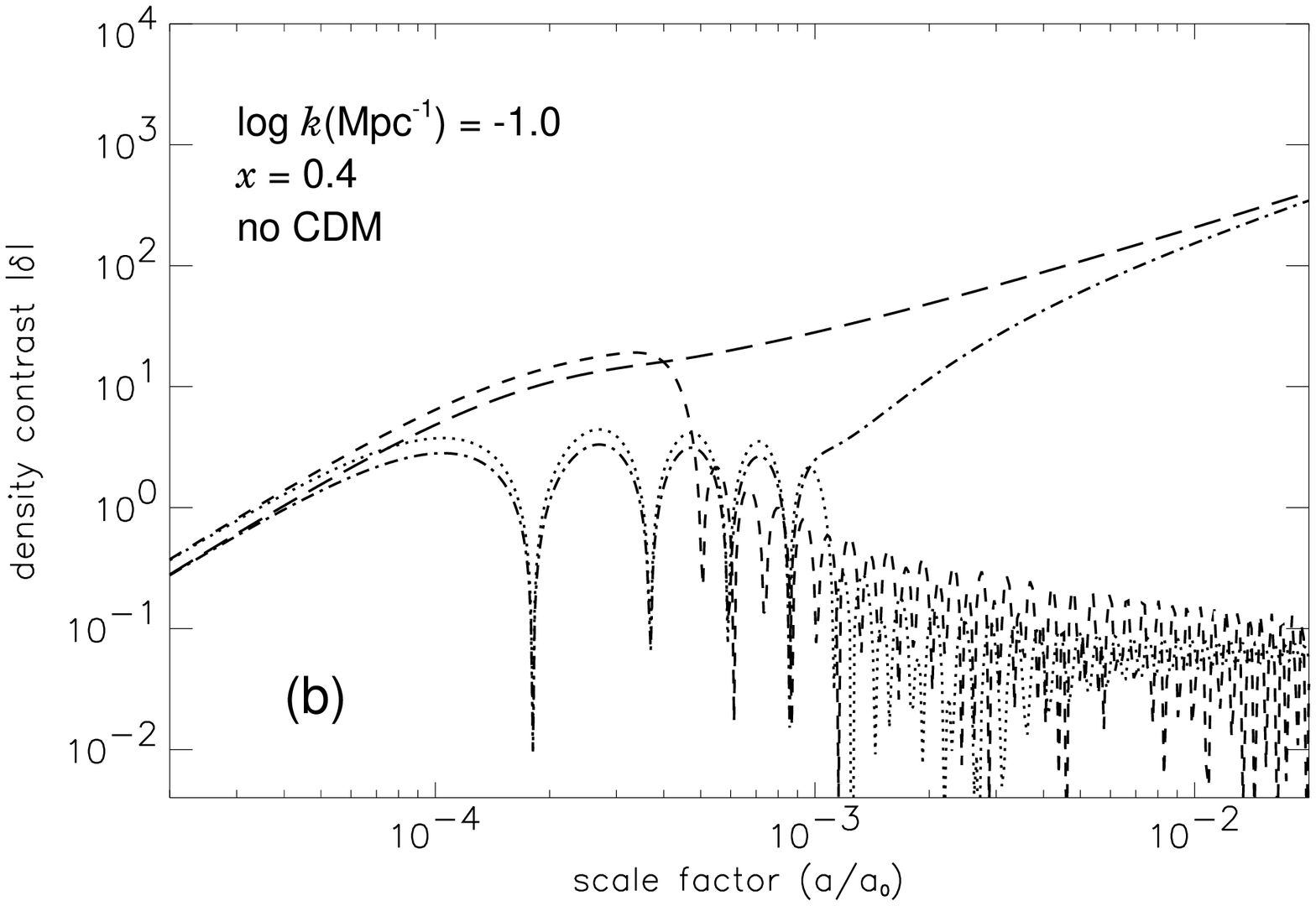} } }
\end{center}
\caption{\small Evolution of perturbations for the components of a Mirror 
Universe: cold dark matter (solid line), ordinary baryons and photons 
(dot-dashed and dotted) and mirror baryons and photons 
(long dashed and dashed). 
The models are flat with $ \Omega_m = 0.3 $, $ \Omega_b h^2 = 0.02 $, 
$ \Omega'_b = 4 \Omega_b $ ($ a $) or $ (\Omega_m - \Omega_b) $ 
(no CDM) ($ b $), $ h = 0.7 $, $ x = 0.4 $, and $ \log k ({\rm Mpc}^{-1}) = -1.0 $.}
\label{evol-x04-b4nocdm-k10}
\end{figure}

Obviously, these are cases where the CDM continues to be the dominant 
form of dark matter, and drives the growth of perturbations, given its 
continuous increase. 
In any case, if the dominant form of dark matter is made of mirror baryons 
the situation is practically the same, as visible comparing figures 
\ref{evol-x0504-b2-k10}b and \ref{evol-x04-b4nocdm-k10}a (where we see 
only slight differences on the CDM and mirror baryons behaviours in the 
central region of the plots), since mirror baryons decouple before than 
ordinary ones and fall into the potential wells of the CDM, reinforcing them.

Finally, in the interesting case where mirror baryons constitute {\em all} the 
dark matter, they drive the evolution of perturbations. 
In fact, in figure \ref{evol-x04-b4nocdm-k10}b we clearly see that the density 
fluctuations start growing in the mirror matter and the visible baryons are 
involved later, after being recombined, when they rewrite the spectrum of 
already developed mirror structures. 
This is another effect of a mirror decoupling occurring earlier than the 
ordinary one: the mirror matter can drive the growth of perturbations in 
ordinary matter and provide the rapid growth soon after recombination 
necessary to take into account of the evolved structures that we see today.

After all the considerations made in this paper, it is evident that the case of 
mirror baryons is very interesting for structure formation, because they are 
collisional between themselves but collisionless for the ordinary sector, or, 
in other words, they are self-collisional. 
In this situation baryons and photons in the mirror sector are tightly coupled 
until decoupling, and structures cannot grow before this time, but the mirror 
decoupling happens before the ordinary one, thus structures have enough 
time to grow according to the limits imposed by CMB and LSS (something 
not possible in a purely ordinary baryonic scenario). 
Another important feature of the mirror dark matter scenario is that, if we 
consider small values of $ x $, the evolution of primordial perturbations is 
very similar to the CDM case, but with a fundamental difference: there exist 
a cutoff scale due to the mirror Silk damping, which kills the small scales, 
overcoming the problems of the CDM scenario with the excessive number 
of small satellites. 
These are important motivations to go further in this work and investigate the 
effects of the mirror sector on the CMB and LSS power spectra, as we will do 
in Paper 2 \cite{paper2}.

\newpage

\section{Conclusions}

We have studied the structure formation in linear regime for a Universe 
with mirror dark matter. 
This topic was introduced in previous works \cite{paolo}, 
and here we analyzed the details of 
the evolution of all the relevant scales (sound speed, Jeans length and 
mass, dissipative scales) comparing with the ordinary and the CDM cases 
and for different mirror scenarios, together 
with the evolution of perturbations varying a lot of parameters.

We extended the Jeans theory to the more general case where the Universe 
is made of two sectors, ordinary and mirror, using a pre-existent bound 
(coming from the BBN limits) on the mirror temperature: in the 
relativistic expansion epoch the cosmological energy density is dominated 
by the ordinary component and the mirror one gives a negligible 
contribution, while for the non-relativistic epoch the complementary situation 
can occur when the mirror baryon matter density is bigger than the ordinary 
one, $ \Omega'_b > \Omega_b $, and hence the mirror baryonic dark matter 
(MBDM) can contribute the dark matter of the Universe along with the CDM 
or even constitute a dominant dark matter component.     

Since the existence of a mirror hidden sector changes the time of 
key epochs, there are important consequences on the structure formation 
scenario for a Mirror Universe. 
We studied this scenario in presence of adiabatic scalar density 
perturbations, which are at present the most probable kind of primordial 
fluctuations, in the context of the Jeans gravitational instability theory.

Given that the physics is the same in both sectors, key differences are the 
shifts of fundamental epochs, namely the baryon-photon 
equipartition and the matter-radiation decoupling 
occur in the mirror sector before than in the ordianary one:
$ a_{\rm b\gamma}' < a_{\rm b\gamma} $; $ a_{\rm dec}' < a_{\rm dec} $. 
The first step was the study of the mirror sound speed and its 
comparison with the ordinary one and with the velocity dispersion of a 
typical cold dark matter candidate. 
From this study we obtained the mirror Jeans length and mass, again to be 
compared with the same quantities obtained for the ordinary sector and for 
the cold dark matter. 
There are two different possibilities, according to the value of $ x $, which 
can be higher or lower than 
$ x_{\rm eq} \approx 0.046 (\Omega_m h^2)^{-1} $, the value for which 
mirror decoupling occurs at matter-radiation equality time. 

The values of the length and mass scales clearly depend on the mirror 
sector temperature and baryonic density, but we found that $ M_{\rm J}' $ is 
always smaller than $ M_{\rm J} $, with a typical ratio $ \sim 50 $ for 
$ x > x_{\rm eq} $, while for cold dark matter it is several orders of 
magnitude lower. For $ x < x_{\rm eq} $, $ M_{\rm J}' $ is few orders 
smaller than the same quantity obtained for the case $ x > x_{\rm eq} $.

Another important quantity to describe the structure evolution is the 
dissipative scale, represented in the mirror sector by the mirror Silk scale. 
We found that it is much lower than the ordinary one, obtaining 
$ M_{\rm S}' \sim 10^{10} \:{\rm M}_\odot $ (around a typical galactic mass) 
for $ x \simeq x_{\rm eq} $, a value similar to the free streaming scale for 
a typical warm dark matter candidate, and much higher than the one for 
cold dark matter.

We collected all these informations in two different mirror scenarios, 
for $ x > x_{\rm eq} $ and $ x < x_{\rm eq} $. 
In the latter case we obtained a maximum mirror Jeans scale similar to the 
mirror Silk mass, so that practically all perturbations with masses greater 
than this Silk mass grow uninterruptedly, just as in a cold dark matter 
scenario.

After this, we modified a numerical code existing for the standard Universe 
in order to take into account a hidden mirror sector, and computed the 
evolution of perturbations in the linear regime for the components 
present in a Mirror Universe, namely the ordinary and mirror baryons and 
photons, and the cold dark matter. 
We did this for various mirror temperatures and baryon densities, and for 
different perturbative scales, finding all the features predicted by our 
structure formation study (as for example the mirror decoupling and the 
CDM-like behaviour for low $ x $-values).

This was the first step in order to obtain the photon and matter power spectra 
for a Universe with mirror dark matter and compare them with observations, 
and this will be exactly the object of the next paper of this series.


\section*{Acknowledgements}

\noindent 
I am grateful to my invaluable collaborators 
Zurab Berezhiani, Denis Comelli and Francesco Villante. 
I would like to thank also 
Silvio Bonometto, Stefano Borgani, Alfonso 
Cavaliere and Nicola Vittorio for interesting discussions.



\end{document}